\documentclass[aps,prd,floatfix,twocolumn,showpacs,superscriptaddress,nofootinbib]{revtex4-1}
\usepackage{color}
\usepackage{graphicx}
\usepackage{aas_macros}
\usepackage{ amssymb }
\usepackage{amsmath}
\usepackage{acronym}
\usepackage{svn-multi}
\usepackage{import}
\usepackage[dvipsnames]{xcolor}

\newcommand{\msun}{\ensuremath{\mathrm{M}_\odot}}
\newcommand{\TheEvent}{GW150914}
\newcommand{\Xmas}{GW151226}

\newcommand{\degs}{\ensuremath{^\circ}}
\newcommand{\chieff}{\ensuremath{\chi_{eff}}}

\newcommand{\WF}{IMRPhenomPv2}
\newcommand{\HLVIJMASSRANGE}{{\ensuremath{[25-100]~\msun}}}
\newcommand{\HLVMASSRANGE}{{\ensuremath{[30-50]~\msun}}}
\newcommand{\si}{\ensuremath{\sim}}
\newcommand{\tjn}{\ensuremath{\theta_{JN}}}
\newcommand{\MONESCOMPACT}{{\ensuremath{36.2_{-3.8}^{+5.2}}}} 
\newcommand{\MTWOSCOMPACT}{{\ensuremath{29.1_{-4.4}^{+3.7}}}} 
\newcommand{\SPINCONE}{{\ensuremath{0.32_{-0.29}^{+0.47}}}} 
\newcommand{\SPINCTWO}{{\ensuremath{0.48_{-0.43}^{+0.47}}}} 
\newcommand{\MONESCOMPACTBoxing}{{\ensuremath{14.2_{-3.7}^{+8.3}}}} 
\newcommand{\MTWOSCOMPACTBoxing}{{\ensuremath{7.5_{-2.3}^{+2.3}}}} 
\newcommand{\sv}[1]{\textcolor{black}{#1}}
\newcommand{\jv}[1]{\textcolor{black}{#1}}

\newcommand{\vivien}[1]{\textcolor{black}{#1}}

\newcommand{\linf}{\texttt{lalinference}}
\begin{document}
\title{Parameter estimation for heavy binary-black holes with networks of second-generation gravitational-wave detectors}
\author{Salvatore Vitale}
\email{salvatore.vitale@ligo.org}
\author{Ryan Lynch}
\affiliation{LIGO, Massachusetts Institute of Technology, Cambridge, Massachusetts 02139, USA}
\author{Vivien Raymond}
\affiliation{Albert-Einstein-Institut, Max-Planck-Institut fur Gravitationsphysik, D-14476 Potsdam-Golm, Germany}
\author{Riccardo Sturani}
\affiliation{International Institute of Physics (IIP), Universidade Federal do Rio Grande do Norte (UFRN) CP 1613, 59078-970 Natal-RN Brazil}
\author{John Veitch}
\affiliation{School of Physics and Astronomy, University of Birmingham, Birmingham, B15 2TT, United Kingdom}
\author{Philip Graff}
\affiliation{Department of Physics \& Joint Space-Science Institute, University of Maryland, College Park, MD 20742, USA}

\begin{abstract}

The era of gravitational-wave astronomy has started with the discovery of the binary black hole coalescences (BBH) GW150914 and GW151226 by the LIGO instruments. These systems allowed for the first direct measurement of masses and spins of black holes. 
The component masses in each of the systems have been estimated with uncertainties of over 10\%, with only weak constraints on the spin magnitude and orientation. In this paper we show how these uncertainties will be typical for this type of source when using advanced detectors. Focusing in particular on heavy BBH of masses similar to GW150914, we find that typical uncertainties in the estimation of the source-frame component masses will be around 40\%.  We also find that for most events the magnitude of the component spins will be
estimated poorly: for only 10\% of the systems the uncertainties in the spin magnitude of the primary (secondary) BH will be below 0.7 (0.8). Conversely, the effective spin along the angular momentum can be estimated more precisely than either spins, with uncertainties below 0.16 for 10\% of the systems.
We also quantify how often large or negligible primary spins can be excluded, and how often the sign of the effective spin can be measured. We show how the angle between the spin and the orbital angular momentum can only seldom be measured with uncertainties below 60$^\circ$. We then investigate how the measurement of spin parameters depends on the inclination angle and the total mass of the source. We find that when precession is present, uncertainties are smaller for systems observed close to edge-on. Contrarily to what happens for low-mass, inspiral dominated, sources, for heavy BBH we find that large spins aligned with the orbital angular momentum can be measured with small uncertainty. 
We also show how spin uncertainties increase with the total mass. Finally, considering a simple toy model, we show how detections can be combined to infer properties of the underlying population.

\end{abstract}
\maketitle

\section{Introduction}

The Advanced LIGO~\cite{Harry:2010zz} observatories have discovered gravitational waves (GWs) emitted by a binary black hole coalescence (BBH), on September 14th 2015~\cite{GW150914-DETECTION}. The event was named~\TheEvent{}. A few months later, a second clear BBH detection (GW151226) was made~\cite{GW151226-DETECTION,O1-BBH}, and a weaker candidate BBH signal (LVT151012) was also identified.

The key astrophysical parameters of these sources have been estimated using Bayesian algorithms~\cite{2015PhRvD..91d2003V,GW150914-PARAMESTIM, GW151226-DETECTION,O1-BBH,2016arXiv160601210T} and tests of general relativity have been performed~\cite{GW150914-TESTOFGR,O1-BBH,2016arXiv160308955Y}. The source-frame masses of the two black holes in \TheEvent{} have been estimated~\cite{GW150914-PARAMESTIM,O1-BBH} to be {\MONESCOMPACT} \msun{} and {\MTWOSCOMPACT} \msun, where the error bars include both statistical and systematic errors from waveform mismatch, with the statistical uncertainty contributing the most. For \Xmas{} the two masses have been estimated with similar large uncertainty to be {\MONESCOMPACTBoxing}~\msun{} and {\MTWOSCOMPACTBoxing}~\msun~\cite{O1-BBH}.

Within general relativity, the dimensionless spin magnitude can take values in the range $[0,1]$, with $0$ being non-spinning and $1$  being maximally spinning. 
For both sources, the spins of the two black holes have been measured with high uncertainty, the 90\% credible interval on the measurement spanning most of the prior support. For \TheEvent{}, the median and 90\% credible interval were {\SPINCONE} and {\SPINCTWO}~\cite{GW150914-PARAMESTIM}.
Something more could be said about the spins of \Xmas{}, for which there was evidence that at least one of the spins was larger than zero~\cite{GW151226-DETECTION,O1-BBH}, but no meaningful constraint on the spin tilt angles has been set, for any of the systems.

Precise estimation of masses and spins of black holes from gravitational-wave
sources will contribute toward the understanding of the formation and the
properties of these objects, and will complement measurements made with
electromagnetic radiation. For example, both masses and spins of black holes can be measured for black
holes in X-ray binaries, but those are indirect measurements. The mass is found
by measuring the mass of the companion object and the projection of the radial velocity along
the line of sight (which is degenerate with the inclination of the orbital
plane)~\cite{Vikhlinin:1999wx}. The masses of several black holes have been
estimated using this method, with values that cover the range $[5-15]$\,\msun~\cite{2010ApJ...725.1918O}.
Two main methods exists to measure
spins~\cite{1997ApJ...482L.155Z,1989MNRAS.238..729F,2003PhR...377..389R,2011CQGra..28k4009M},
both of which rely on modeling the disk surrounding the BH. 
The mass and spin estimation  of \TheEvent{} and \Xmas{} thus represent the first direct measurements of such quantities.

The main astrophysical implications of the discoveries have been discussed in
\cite{GW150914-ASTRO,O1-BBH}, while a prediction of the rate of heavy BBH
coalescence and prospects for detection in future observing runs was given in~\cite{GW150914-RATES}
(and later updated in \cite{O1-BBH}). The rate estimates suggest that the number
of significant BBH detections by ground based detectors could already be around one per month
in the second observing run, starting before the end of 2016~\cite{2016LRR....19....1A}.
In view of the numerous detections that will be made in the next few years,
it is worth addressing the following question: was the precision in the
measurement of parameters for the detected systems typical of what we
can expect in the future?
\sv{In this paper we address that question. 
Since results already exist in the literature (see below) for lighter BBH, here we will thus focus on heavy BBH. These are systems that will only be in band for a few cycles before merger, thus making unclear a priori, what and to which precision can be deduced about the individual binary constituents parameters.
Furthermore, given that advanced detectors have a selection bias toward higher masses~\cite{2016PhRvD..94l1501V}, one might expect heavy BBH to be detected more often, if the rates are comparable to those for stellar-mass BBHs.}

\jv{Some previous studies of parameter estimation for BBH (including heavy BBH) have been performed. \cite{Ghosh:2015jra}} \sv{considers BBHs with spins aligned with the orbital angular momentum (i.e. without spin-induced precession) and reports statistical uncertainties for the main astrophysical parameters. Their results are comparable to ours for BBH of similar mass, although \jv{the reported uncertainties are} slightly smaller since they don't have potential correlations coming from the precessing spin degrees of freedom. More recently, \jv{\cite{Pankow:2016udj}} looked at neutron star - black hole systems. Their uncertainties in the spin parameters are smaller than what we find here, consistently with the fact that larger mass ratios enhance the measurability of spins~\cite{PhysRevLett.112.251101}. Most of the early work, e.g. \cite{2009CQGra..26k4007R,2008ApJ...688L..61V}, deals with only a few systems at the time, using post-newtonian inspiral-only waveforms. As such, these papers are not directly comparable to ours.}

We create an astrophysical population of 200 spinning heavy BBHs and estimate
their parameters with a network of advanced LIGO and Virgo detectors at design
sensitivity. We find that source-frame component masses can be estimated with
typical uncertainties of 40\%. This is slightly larger than what was measured
for \TheEvent, owing to its large signal-to-noise ratio.
Spin magnitude is hard to estimate: for the most (least) massive black-hole
in the system we find that only 10\% of the times the 90\% credible interval
uncertainty will be smaller than 0.7 (0.8). Similar conclusions hold for the
tilt angles, i.e. the angle between each spin vector and the orbital angular
momentum, for which the uncertainties will be larger 60\degs{} for most systems.
As we mentioned above, \TheEvent{} fits perfectly in this scenario.
We quantify how often large and negligible spins can be excluded, and find
that large spins are easier to exclude. For example, if only BBH with primary
spins up to 0.2 are considered, 90\% of the times spins above 0.95 can be excluded.
We also verify that the effective spin along the orbital angular momentum~\cite{GW150914-PARAMESTIM,O1-BBH} can be estimated more precisely than the individual spins, and that 70\% of the times one can correctly measure the sign of the effective spins, if the underlying population has even mild effective spins (below -0.3 or above 0.3).

We then show how precessing spins can be estimated more precisely as the inclination angle moves away from zero, and the uncertainties reach a minimum for angles close to $\pi/2$, where the binary is viewed in alignment with the orbital plane.
Contrarily to what expected for low mass sources, dominated by the inspiral phase, we find that for heavy BBH spins aligned with the orbital angular momentum are not extremely degenerate with the mass parameters, and can thus be measured very precisely. In fact, considering BBH with mass ratios of 1 and 2 and spin magnitude of 0.9, we find that aligned spins can be measured a factor of several better than precessing spins, no matter of the orbital orientation.

We  investigate how the uncertainties depend on the (redshifted) total mass of the system and find that the uncertainties increase with the total mass, with larger increases for larger mass ratios.

Finally, we show how the properties of the underlying astrophysical distribution can be estimated, in a very simple toy model.

The rest of this paper is organized as follows. We first describe the GW network (Sec.~\ref{Sec.Detectors}) and BBH events (Sec.~\ref{Sec.Events}) used in this study.
The main results are summarized in Sec.~\ref{Sec.Results}, while conclusions and discussion can be found in Sec.~\ref{Sec.Conclusions}.

\section{Method}
\subsection{Detectors}\label{Sec.Detectors}

In this study, we consider a network of 3 advanced detectors, the two LIGO interferometers (IFOs) and the Virgo detector. (In Appendix~\ref{App.HLVIJ} we will consider a five-detector network includes the three instruments above plus the KAGRA detector in Japan~\cite{2012CQGra..29l4007S} and LIGO India~\cite{Indigo}.)
For all instruments, we used the noise spectral density corresponding to their design sensitivities~\cite{Harry:2010zz,TheVirgo:2014hva}. We are thus focusing on instruments that will be available later in the decade.  However it is easy to realize that the main results we obtain will not strongly depend on this choice. The main difference on the detected events if the instruments are made more sensitive is that the distance distribution of the detected events will get shifted to higher values, while keeping roughly the same shape~\footnote{\label{Foot.Caveat} This would not be true if the increase in sensitivity is such that sources at redshifts of several are reached~\cite{Vitale3G,VitaleDiff}, however this won't be the case for systems in the mass range we consider.}. Critically, the distribution of the signal-to-noise ratios (SNRs)  will be the same. Since the uncertainty in the intrinsic parameters (mass, spins) mostly depend on the SNR, with the caveat above~$^{\ref{Foot.Caveat}}$, the distribution of uncertainties we obtain should be representative of the uncertainties of the next few years. In fact, we will see that the uncertainties of \TheEvent{} follow very well the ones we obtain here.

\subsection{Simulated events}\label{Sec.Events}

\begin{figure}[htb]
\includegraphics[width=0.98\columnwidth]{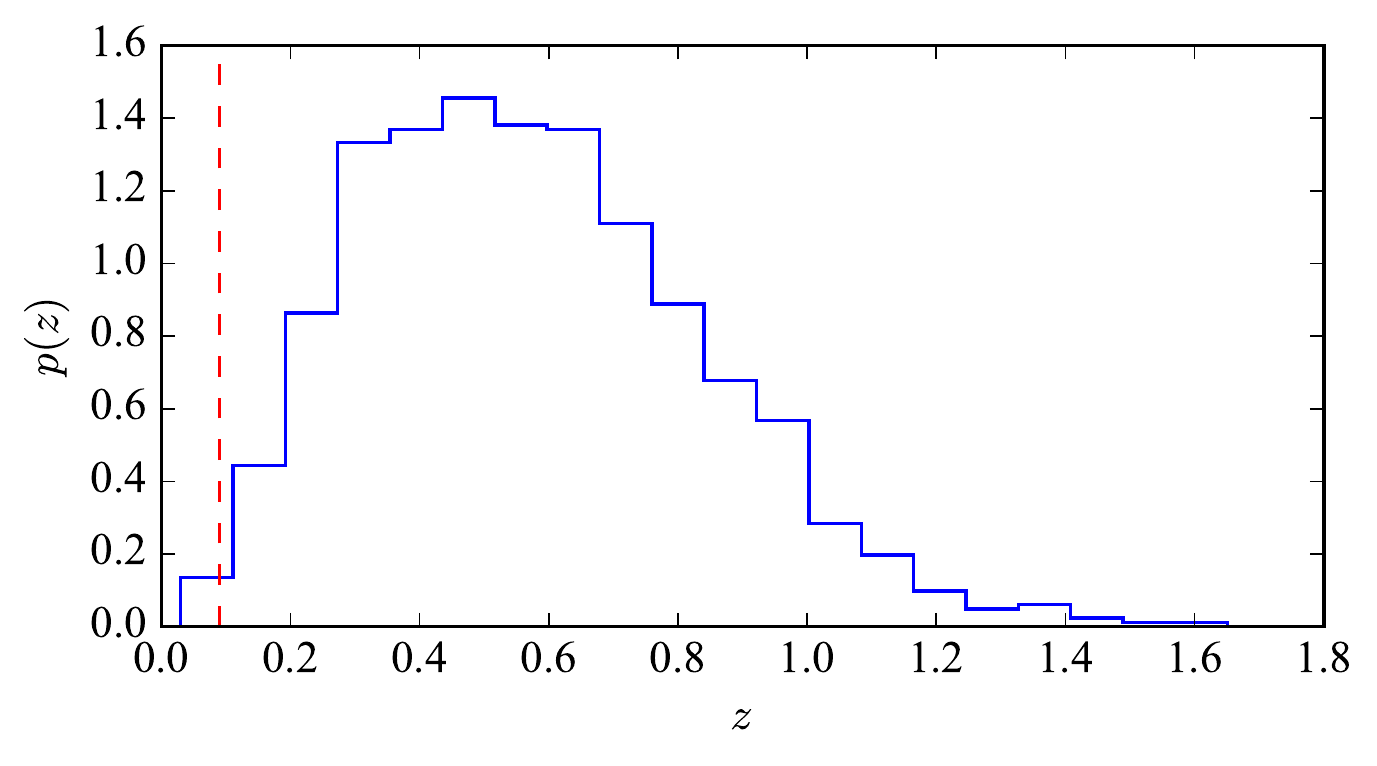}
\caption{The redshift distribution of detectable heavy BBH with a network of
LIGO and Virgo at design sensitivity. The vertical dashed line is the median estimated redshift for \TheEvent~\cite{O1-BBH}}\label{Fig.InjRedshift}
\end{figure}

We simulated $200$ binary black hole systems with intrinsic masses uniformly
        drawn from the range \HLVMASSRANGE, and dimensionless spin magnitudes,
$a\equiv \frac{c|\vec{ S}|}{G m^2}$, drawn uniformly from the range
$[0,0.98]$.
The sky position and orientation of the systems are isotropically distributed.
The distances are drawn uniformly in comoving volume, with a lower network signal-to-noise
ratio (that is, the quadrature sum of the SNR in each instrument) cut at
$12$. This corresponds to distances up to \si12~Gpc, or a redshift of $\sim 1.5$ using a
$\Lambda$CDM flat cosmology~\cite{2015arXiv150201589P}.
The redshift distribution of the simulated signals is shown in Fig.~\ref{Fig.InjRedshift}
with a vertical line showing the median measured redshift of \TheEvent{}. 

We notice that \TheEvent{} is on the left side of the distribution since it was detected by 2 LIGO
instruments at early sensitivity~\cite{GW150914-PARAMESTIM,GW150914-DETECTORS},
while in this paper we consider a network made of more sensitive detectors.

\begin{figure}[htb]
\includegraphics[width=0.98\columnwidth]{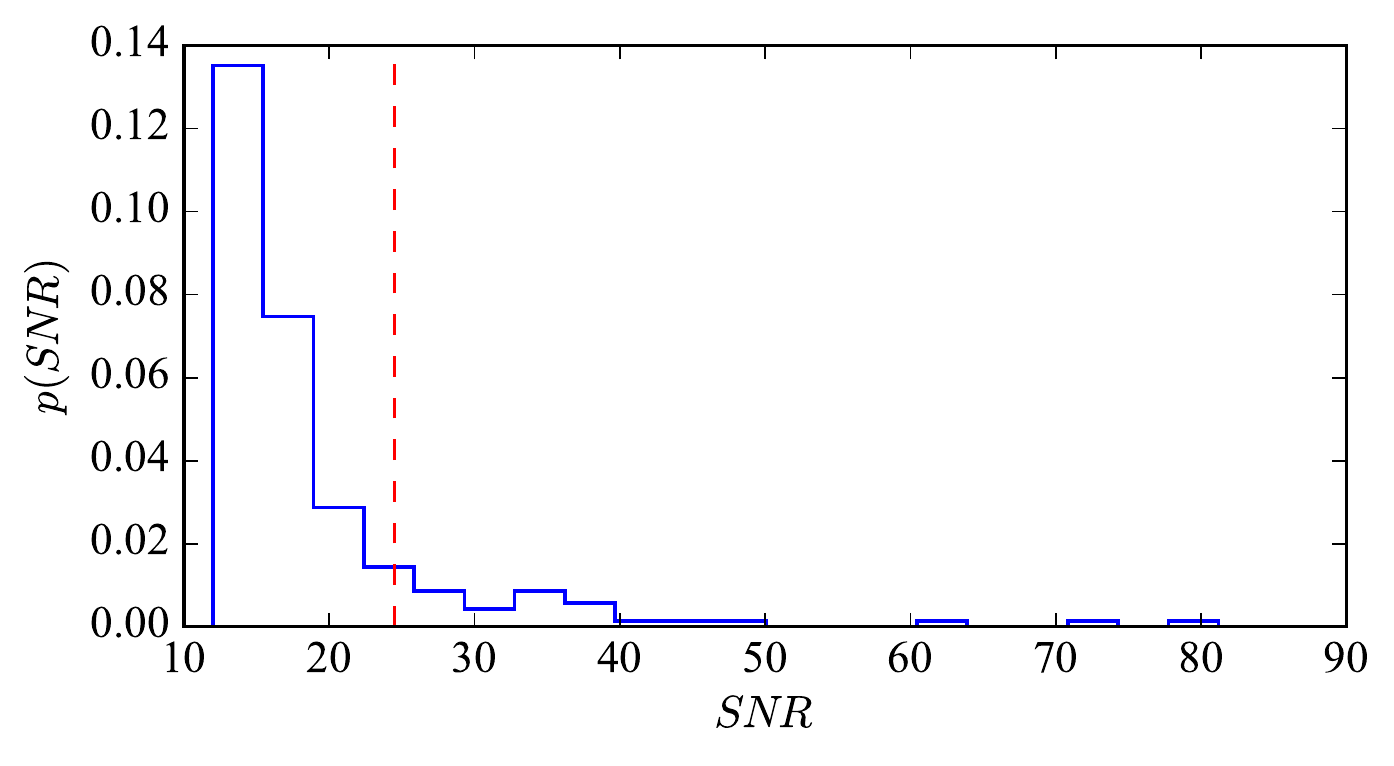}
\caption{The SNR distribution of detectable heavy BBH with a network of
LIGO and Virgo at design sensitivity. The vertical dashed line is the SNR of \TheEvent~\cite{O1-BBH} }\label{Fig.InjSNR}
\end{figure}

\sv{In Fig.~\ref{Fig.InjSNR} we show the network SNR of the population of BBH. Here too, a vertical dashed line shows the SNR of \TheEvent{}. We see that, even when considering a 3-detector network at design sensitivity, as we do in this work, \TheEvent{} is considerably louder than the ``typical'' detection.}

The simulations were performed using the IMRPhenomPv2
waveform approximant~\cite{Hannam:2013oca,2016PhRvD..93d4007K,2016PhRvD..93d4006H}.
This is a phenomenological inspiral-merger-ringdown approximant, and is one of the
two used to estimate the parameters of the detected events~\cite{GW150914-PARAMESTIM,GW151226-DETECTION,O1-BBH}.
It must be stressed that \WF{} uses a simplified spin description~\cite{Hannam:2013oca,2015PhRvD..91b4043S},
in which the main spin parameters are the effective component of the total spins
along the orbital angular momentum ($\chi_\mathrm{eff}$ in ~\cite{GW150914-PARAMESTIM}) and perpendicular to it ($\chi_p$ in \cite{GW150914-PARAMESTIM}). The magnitude and orientations of the component spins can be obtained from those. Although \WF{} uses a simplified spin prescription, it has been shown for \TheEvent{} that the results obtained with \WF{} broadly agree with those obtained with a fully precessing time-domain approximant (SEOBNRv3)~\cite{2016arXiv160601210T}. The agreement might be inferior in some corners of the parameter space (e.g. for systems seen from edge-on, i.e. with their orbital angular momentum forming an angle of $\iota\sim \pi/2$ with the line of sight)~\cite{PuerrerPrep}. However, \WF{} is more than one order of magnitude faster to compute than SEOBNRv3 for the masses considered in this study. Considering that a single parameter estimation run requires the computation of $\sim 10^6$ waveforms, we will thus work with the former. 
When surrogate Reduced Order Models (ROMs)~\cite{2016PhRvD..93f4041P,2014CQGra..31s5010P,2014PhRvL.113b1101B,2014PhRvX...4c1006F} become available for SEOBNRv3, possibly followed-up by Reduced Order Quadrature~\cite{2015PhRvL.114g1104C,Smith:2016qas}, this study should be repeated. However the main conclusions of this study should hold, since most events detected by advanced detectors will be oriented close to face-on ($\iota\sim0$) or face-off ($\iota\sim\pi$)~\cite{2011CQGra..28l5023S}.

By using the same waveform family to simulate the signals and to estimate their parameters, we do not consider any effect of waveform systematics.
In practice different waveform families will always lead to slightly different parameter estimates, but here we assume that those difference will keep becoming smaller in the next months and years, as more and more elaborate waveform families are introduced.

All simulated BBH are added (``injected") into simulated interferometric data of advanced LIGO and Virgo.

Algorithms to estimate parameters of spinning CBC have been developed over the last several years, based on either Monte Carlo~\cite{2008ApJ...688L..61V,2009CQGra..26t4010V} or nested sampling~\cite{2010PhRvD..81f2003V} methods.
In this paper we use the algorithm that yielded estimates for the two detected events~\cite{O1-BBH}, \texttt{lalinference}~\cite{2015PhRvD..91d2003V}.

\section{Results}\label{Sec.Results}

In what follows we will use the symbol $\Sigma_x$ to refer to the 90\% credible interval (CI) for the parameter x (with dimensions), and the symbol $\Gamma_x$ for the relative uncertainty w.r.t. the true value: $\Gamma_x\equiv \Sigma_x / x_\mathrm{true}$ (dimensionless).
Our $\Sigma$ will thus be directly comparable with the uncertainties as reported for \TheEvent{} and \Xmas.

\subsection{Masses}\label{Sec.Masses}

We start by looking at the estimation of mass parameters. 
When sources at non-negligible redshifts are being detected, one must distinguish between the intrinsic (or source-frame) masses, and the detector-frame (or redshifted) masses.
Using an index $s$ for source-frame quantities and an index $d$ for detector-frame quantities, the relationship is trivially:
\begin{equation}
m^d= (1+z) m^s
\end{equation}
where here with $m$ we generically indicate any 
mass parameter.
In what follows we will use $m_i$ for the component masses, $M$ for the total mass and $q=m_2/m_1 \in [0,1]$
for the asymmetric mass ratio. All masses will be expressed in units of solar masses.
We will examine both intrinsic and redshifted masses, because while intrinsic masses are what is astrophysically relevant it is the redshifted masses that control the shape and phase evolution of the signals in the instruments and hence impact the uncertainties.

It is known that for low mass CBC such as binary neutron stars or systems containing stellar-mass BHs, GWs can yield extremely precise measurement of the chirp mass $\mathcal{M}\equiv \frac{(m_1\,m_2)^{3/5}}{M^{1/5}}$. On the other hand, uncertainties are larger for the measurement of the component masses, total mass and mass ratio~\cite{2014ApJ...784..119R,2014PhRvD..89b2002V,2016ApJ...825..116F}.
This happens because the chirp mass enters the waveform phase at the lowest order in the inspiral,
while the mass ratio (and thus the component masses and total mass) enters at higher orders (see e.g. \cite{2005PhRvD..71h4008A}).
The situation is different for the heavy BBH we consider in this work, since not only the inspiral, but also the merger and ringdown phases will be in the bandwidth of the detectors. Since those depend on the \emph{total} mass, we can expect similar uncertainties for the chirp mass and the total mass~\cite{2015PhRvL.115n1101V,2015PhRvD..92b2002G,2016MNRAS.457.4499H}.
Furthermore, since the length of the inspiral phase shortens as the mass increases~\cite{GW151226-DETECTION,O1-BBH}, the measurement of the chirp mass should slightly worsen as the masses increase.

\begin{figure}[htb]
\includegraphics[width=0.99\columnwidth]{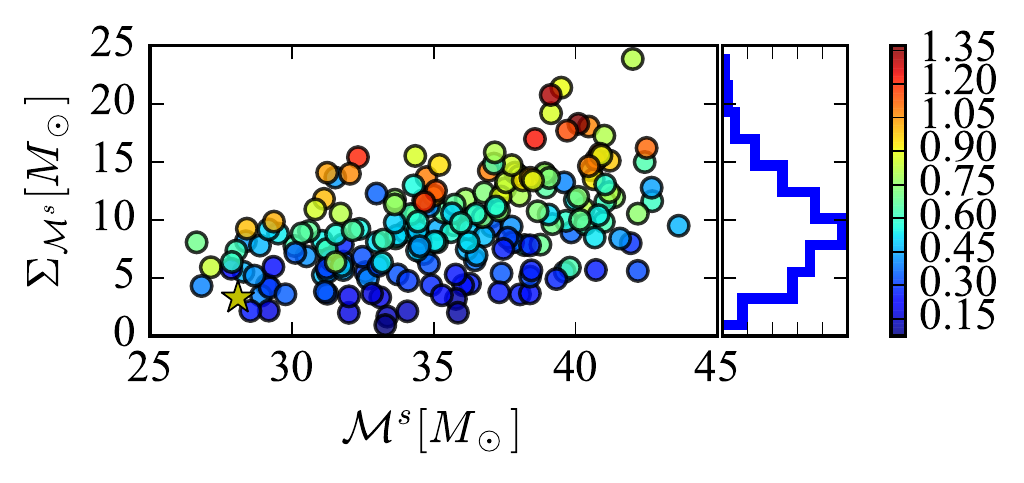}
\caption{The distribution of the 90\% CI uncertainty in the estimation of the source-frame chirp mass (y axis) against the true source-frame chirp mass (x axis). The colorbar is the redshift of the sources. A star reports the coordinates of \TheEvent{}. }\label{Fig.SigmaMCS_MCS}
\end{figure}

In Fig.~\ref{Fig.SigmaMCS_MCS} we report the 90\% CI for the source-frame chirp mass measurement (y axis) against the true injected source-frame chirp mass, while the colorbar reports the injected redshift $z$. Here and in other plots (unless otherwise indicated) a yellow star reports the values for GW150914 (since we don't know the ``true'' value in this case, the x axis refers to the median measured values as given in \cite{O1-BBH}). 

\begin{figure}[htb]
\includegraphics[width=0.99\columnwidth]{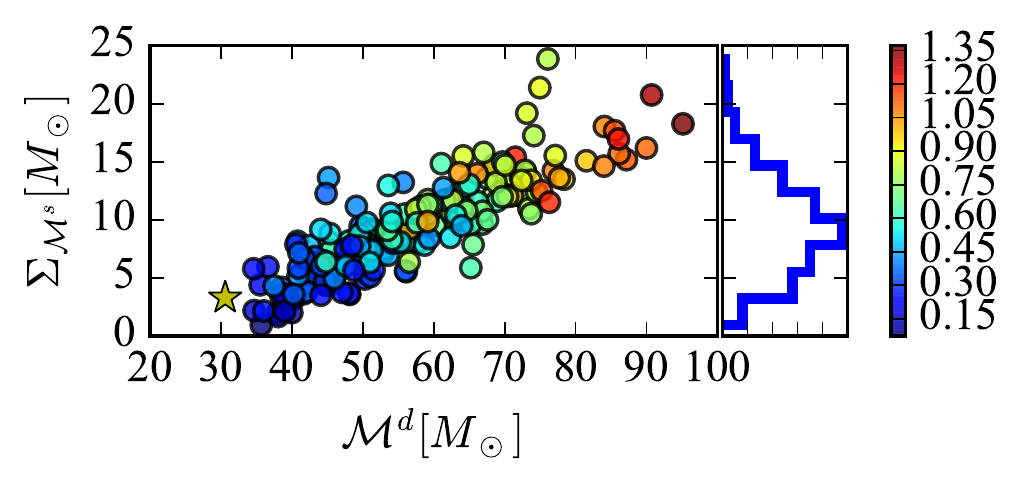}
\caption{The distribution of the 90\% CI uncertainty in the estimation of the source-frame chirp mass (y axis) against the true detector-frame chirp mass (x axis). The colorbar is the redshift of the sources. A star reports the coordinates of \TheEvent{}. }\label{Fig.SigmaMCS_MCD}
\end{figure}

We do not see a strong correlation between injected mass and uncertainties. The only clear trend is that closer events have smaller uncertainties, due to their high SNRs. 
What is happening is that, as mentioned above, the shape of the signal in the detector will depend on the detector-frame masses, and thus on the redshift. If one plots the uncertainties against the detector-frame chirp mass, Fig.~\ref{Fig.SigmaMCS_MCD}, then the correlation becomes evident.

\begin{figure}[htb]
\includegraphics[width=0.99\columnwidth]{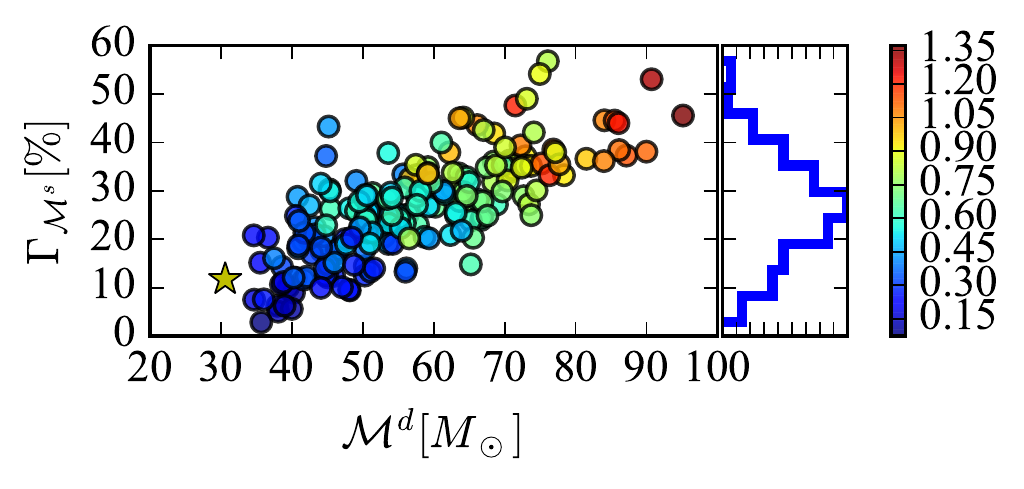}
\caption{The distribution of the 90\% CI relative uncertainty (in percent over the true value) in the estimation of the source-frame chirp mass (y axis) against the true detector-frame chirp mass (x axis). The colorbar is the redshift of the sources. A star reports the coordinates of \TheEvent{}. }
\label{Fig.GammaMCS_MCD}
\end{figure}

Typical uncertainties span a broad range, from a few to $\sim20$\msun, depending on the \emph{detector frame} chirp mass.
This translates to relative uncertainties (over the injected value) in the range few-$60\%$, as shown in Fig.~\ref{Fig.GammaMCS_MCD}, where once again the colorbar reports the redshift, with a peak at \si$30\%$.

In all these plots, we see that the uncertainties for \TheEvent{} seem to be quite typical of systems with comparable masses. 


From an astrophysical point of view, the most relevant mass parameters are the component masses and, relatedly, the mass ratio.
In fact, measuring the masses of heavy BH would allow to estimate their mass distributions, while the mass ratio can be used to distinguish formation channels~\cite{2016PhRvD..93h4029R}.

\begin{figure}[htb]
\includegraphics[width=0.99\columnwidth]{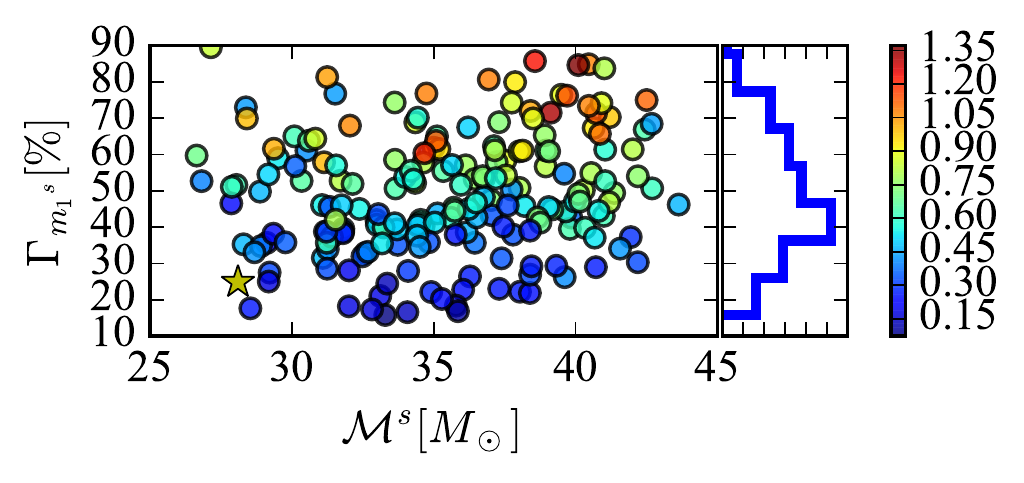}
\caption{The distribution of the 90\% CI relative uncertainty (in percent over the true value) in the estimation of the source-frame primary mass (y axis) against the true source-frame chirp mass (x axis). The colorbar is the redshift of the sources. A star reports the coordinates of \TheEvent{}.}
\label{Fig.GammaM1S_MCS}
\end{figure}

In Fig.~\ref{Fig.GammaM1S_MCS} we show the relative uncertainties for the source-frame mass of the primary BH (i.e. the most massive) against the intrinsic chirp mass. 
We see that 90\% CI uncertainties of the order of several tens of percent will be common for quiet events, while nearby or loud events can have uncertainties of a few tens of percent. \TheEvent{} lives near the tail of the distribution, with uncertainty of $\sim25\%$, since its SNR was large ($\sim 23.7$).
The histogram on the right side reports the distribution of the uncertainties. For a population like the one we considered here, the peak is at $\sim 40\%$.

\begin{figure}[htb]
\includegraphics[width=0.99\columnwidth]{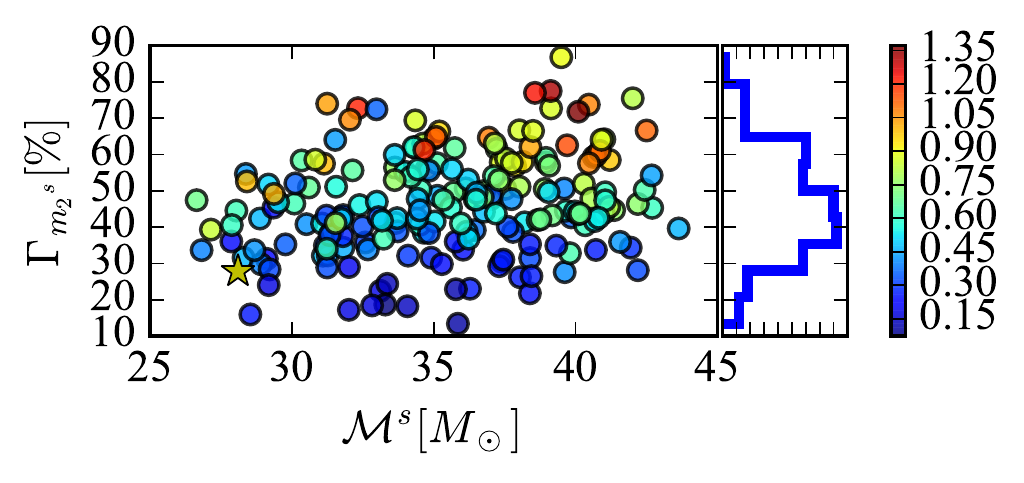}
\caption{Like Fig.~\ref{Fig.GammaM1S_MCS}, but for the secondary BH mass. The colorbar is the redshift of the sources.}
\label{Fig.GammaM2S_MCS}
\end{figure}

A similar plot for the secondary object is shown on Fig.~\ref{Fig.GammaM2S_MCS}. We see that the uncertainties are similar to what obtained for $m_1$, with a slightly larger median.

\begin{figure}[htb]
\includegraphics[width=0.99\columnwidth]{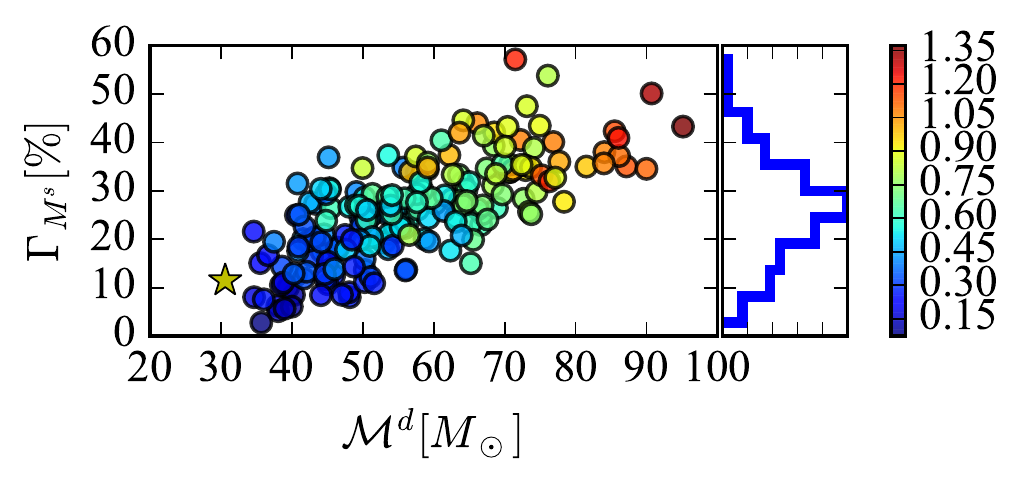}
\caption{The distribution of the 90\% CI relative uncertainty (in percent over the true value) in the estimation of the source-frame total mass (y axis) against the true detector-frame chirp mass (x axis). The colorbar is the redshift of the sources. A star reports the coordinates of \TheEvent{}.}
\label{Fig.GammaMTS_MC}
\end{figure}

Earlier in this section we mentioned that for heavy BBH, we expect the total mass to be estimated as well as the chirp mass (while for BBH of hundreds of solar masses, it will be estimated better than the chirp mass~\cite{2015PhRvL.115n1101V,2015PhRvD..92b2002G,2016MNRAS.457.4499H}). This is indeed confirmed by Fig.~\ref{Fig.GammaMTS_MC}, where we see that typical uncertainties in the measurement of the source-frame total mass will be of a few tens of percent, with a peak of probability at $\sim25\%$. 

\subsection{Spins}

The uncertainties for the \jv{spin magnitudes} for \TheEvent{} covered most of the prior range, with only extreme spins excluded~\cite{GW150914-PARAMESTIM,O1-BBH}. In \cite{PhysRevLett.112.251101} we have shown how uncertainties will generally be large for systems with comparable masses, unless the systems are observed from to edge-on. However, in that paper we only considered a few corners in the parameter space, and worked with stellar mass black-holes.
In this section we wish to show what spin estimation will look like for an astrophysical distribution of more massive BBH.

\begin{figure}[htb]
\includegraphics[width=0.99\columnwidth]{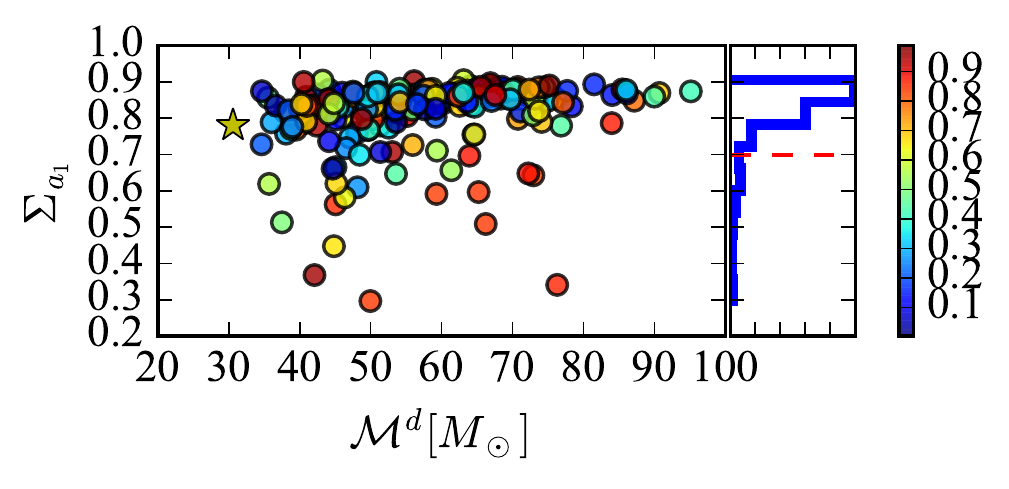}
\caption{The distribution of the 90\% CI uncertainty in the estimation of the primary spin magnitude (y axis) against the true detector-frame chirp mass (x axis). The colorbar shows the magnitude of the primary spin. A star reports the coordinates of \TheEvent{}. The dashed line on the histogram marks the abscissa of the $10^{th}$ percentile.}
\label{Fig.SigmaA1_MC}
\end{figure}

In Fig.~\ref{Fig.SigmaA1_MC} we show the 90\% CI uncertainty in the measurement of the spin magnitude for the most massive BH (y axis) as a function of the redshifted chirp mass. The true spin magnitude is reported in the colorbar. The histogram on the right shows the distribution of the uncertainties. 

We find that larger spins are often easier to measure while for small spins only occasionally the 90\% CI does not cover 90\% of the prior.

The dashed red line in the right panel shows the position of the 10th percentile of the uncertainty distribution, at $\Sigma_{a_1}=0.7$. We thus expect that only in 10\% of the cases we will be able to measure the spin magnitude of the primary BH with an uncertainty smaller than 0.7. We do not see a clear correlation of spin uncertainties with the redshifted chirp mass since too many other factors affect the measurability of the spins. Later, in Sec.~\ref{Sec.DepOnMass} we will investigate how the spin measurement depends on the mass, working with a controlled setup.

We have indicated with a yellow star the median recovered spin magnitude and the uncertainty for GW150914, which we see is totally consistent with the uncertainty of the BBH we simulated.

\begin{figure}[htb]
\includegraphics[width=0.99\columnwidth]{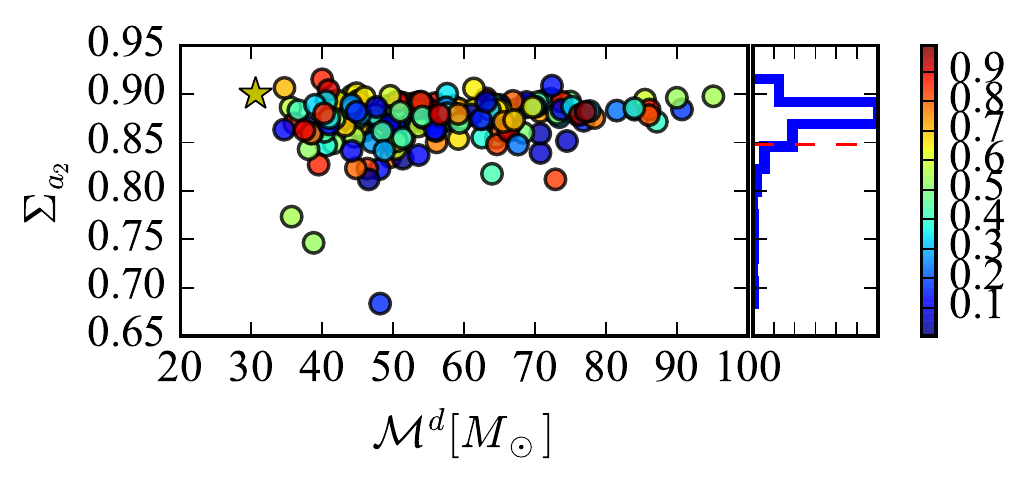}
\caption{Same as Fig.~\ref{Fig.SigmaA1_MC}, but for the secondary spin magnitude. The colorbar shows the magnitude of the secondary spin. Note the different range in the y axis.}
\label{Fig.SigmaA2_MC}
\end{figure}

The same type of plot but for the secondary spin is shown in Fig.~\ref{Fig.SigmaA2_MC} (note the different range in the y axis). As expected, the uncertainties are much larger for the secondary object (10th percentile at 0.85). We thus conclude that it be extremely hard to measure the spin magnitude of the secondary object in heavy BBH systems. This conclusion was reached by \cite{PhysRevD.93.084042} for spin-aligned BBH, and by \cite{PhysRevLett.112.251101} for a few precessing stellar mass BBHs.

Two spin values which have special meaning are obviously zero and one, i.e. no spinning and maximally spinning. In fact, one of the main conclusions of the \TheEvent{} analysis is that the primary BH was not maximally spinning~\cite{GW150914-PARAMESTIM}, whereas for \Xmas{} zero-spin for at least one of the BHs was excluded with high confidence~\cite{GW151226-DETECTION,O1-BBH}.

We have used subsets of our BBH to verify how often we will be able to exclude the extreme scenarios of non spinning and maximally spinning.
We will focus on the primary spin since, as we just saw, the secondary is hardly ever measurable.

\begin{figure}[htb]
\includegraphics[width=0.95\columnwidth]{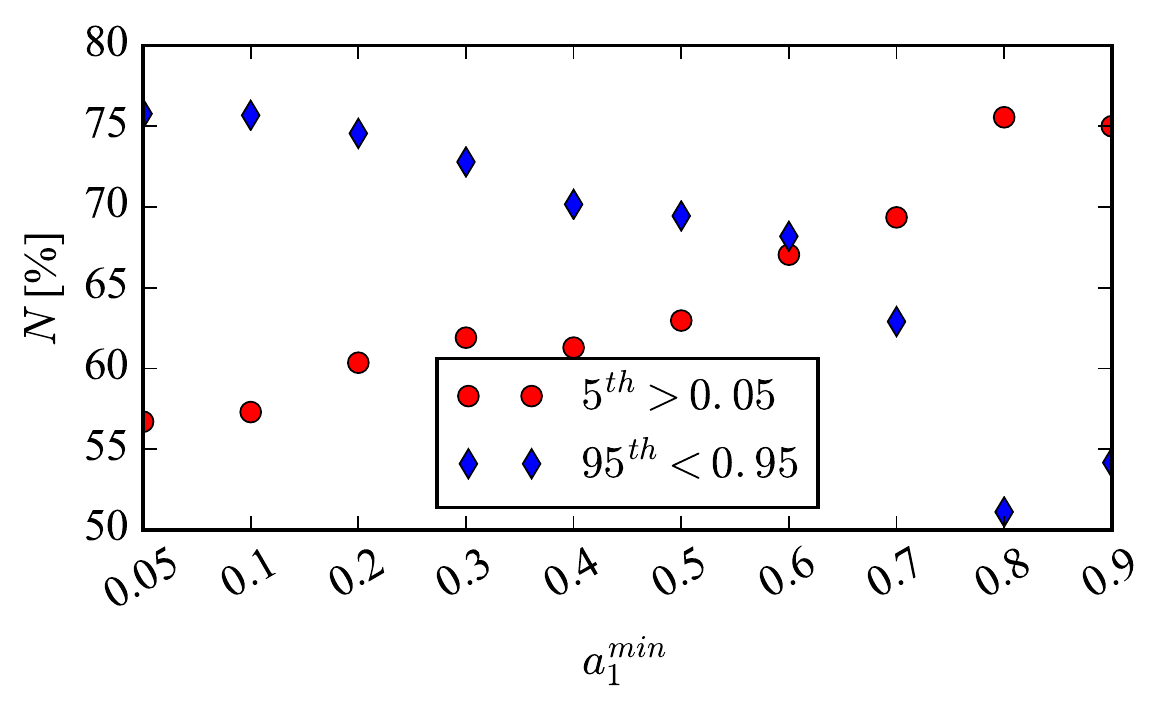}
\caption{In the x axis we give the minimum value of the spin magnitude of the primary BH. The red circles give the fraction of events (y axis) with that minimum spin for which the $5^{th}$ posterior percentile is larger than 0.05. The blue diamonds report the fraction of events for which the $95^{th}$ percentile is smaller than 0.95. If the underlying population is made of BH with large spins (right side of the plot) $\sim75\%$ of the times one can exclude that the primary BH had negligible spin. }
\label{Fig.SpinMin}
\end{figure}

Let us first check what conclusions we can draw from sub-populations of BBH with increasingly large primary spins. 
From our set of BBH we downselect events with increasingly large minimum values of $a_1$, from 0.05 to 0.90. We then check for which fraction of them we can exclude no-spinning and maximally-spinning BHs.
This is shown in Fig.~\ref{Fig.SpinMin}. The left of the plot, with $a_1^{min}=0.05$, thus corresponds to assuming that the astrophysical distribution of $a_1$ is flat in most of the allowed spin range. At the other extreme, on the right of the plot one is assuming that nature only produces BHs with large spins in CBCs. 
Let us first focus on the red circles. They report the fraction of BBH having minimum spin magnitude given in the x axis for which one can conclude $a_1>0.05$ at 90\% CI.
As one would expect, the worst result is obtained when we keep nearly all events ($a_1^{min}=0.05$) since that will include events with small spins, for which it will be hard to exclude low spin values (or actually, to draw any conclusion). As we increase the minimum value of the true spin magnitude, moving to the right of the plot, the fraction of events for which we can exclude small spins increases until it reaches $\sim 75\%$ when we only keep sources with large spins.
We remark that this fraction doesn't go close to 100\%, and even when all systems have large spin on the primary, for $\sim20$\% of them we won't be able to exclude the absence of spin.
The blue diamonds in the same plot quantify the fraction of events for which we can exclude than $a_1$ is larger than 0.95, again at the 90\% CI.
The curve is roughly a mirror of the previous one. If a whole distribution of spin is considered ($a_1^{min}=0.05$), roughly 75\% percent of the time one can exclude very large spins. As the spins increase in the underlying population, the efficiency goes naturally down, until it reaches $\sim50$\%. 

One might be surprised that even when the minimum spin is large (say 0.9) it is still the case that $\sim50\%$ of the times the 95th percentile is smaller than 0.95. This does happen because for most events, not matter of their spins, the posterior distribution for $a_1$ will be centered in the middle of the prior, with errorbars that cover a large fraction of the prior (see Fig.~\ref{Fig.SpinScatter} below and the related discussion).

\begin{figure}[htb]
\includegraphics[width=0.95\columnwidth]{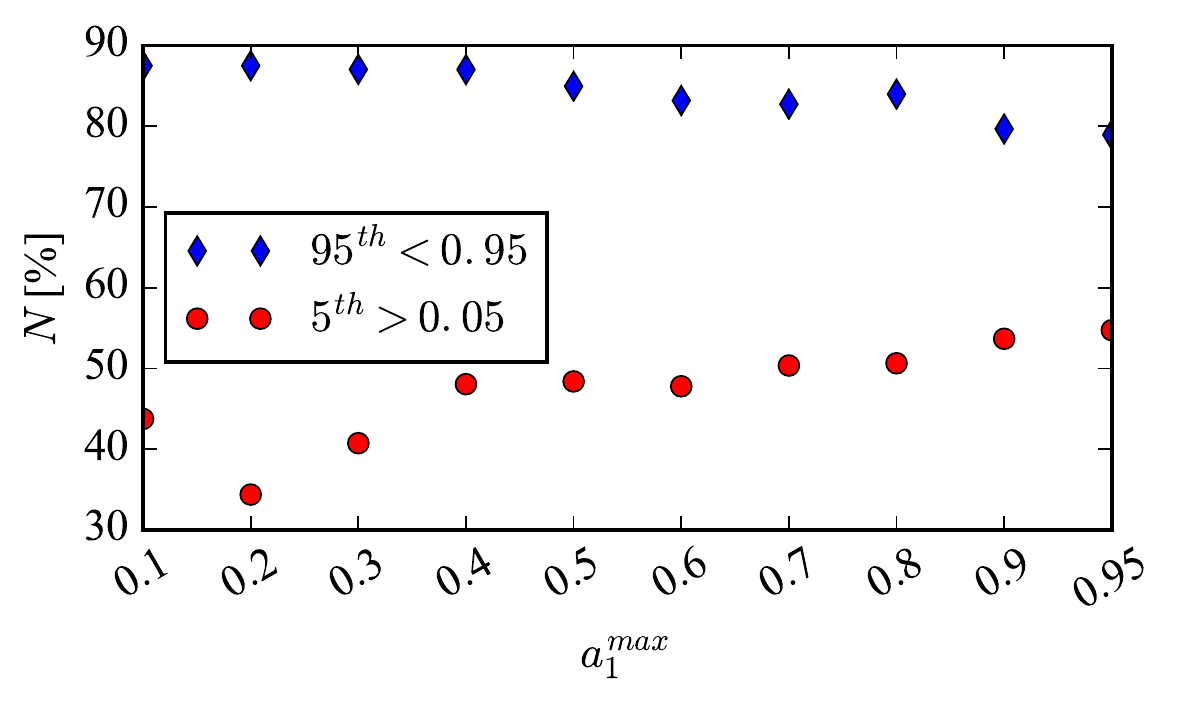}
\caption{In the x axis we give the maximum value of the spin magnitude of the primary BH. The red circles give the fraction of events (y axis) with that maximum spin for which the $5^{th}$ posterior percentile is larger than 0.05. The blue diamonds report the fraction of events for which the $95^{th}$ percentile is smaller than 0.95.  If the underlying population is made of BH with small spins (left side of the plot) $\sim90\%$ of the times one can exclude that the primary BH had large spin.}
\label{Fig.SpinMax}
\end{figure}

\begin{figure}[htb]
\includegraphics[width=0.95\columnwidth]{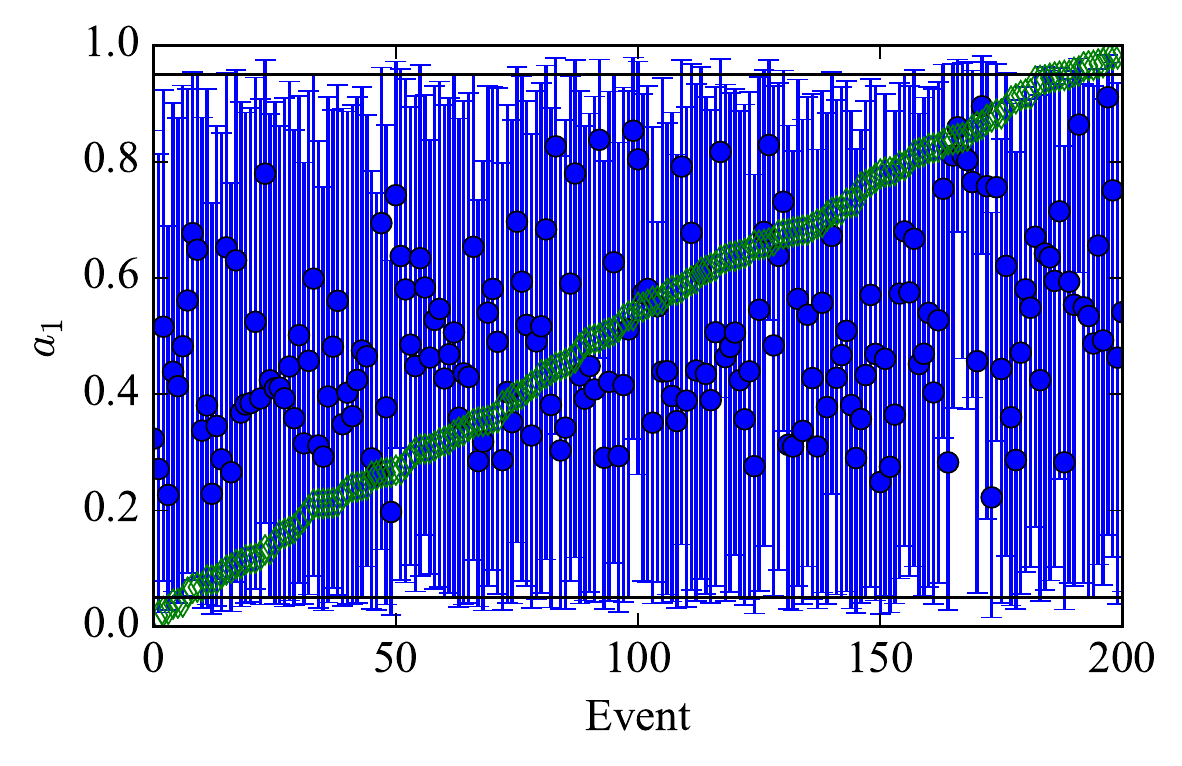}
\caption{For all events, median value of the primary BH spin with 90\% CI. The empty green diamonds indicate the true spins. The two horizontal black lines are at 0.05 and 0.95.}
\label{Fig.SpinScatter}
\end{figure}

We next perform the opposite exercise, and downselect events with decreasing maximum primary spin, given in the x axis of Fig.~\ref{Fig.SpinMax}. 
Once again, the red circles report the fraction of events for which negligible primary spin can be excluded. We see that this fraction is nearly always below 0.5. 
Looking at the blue diamonds, i.e. the fraction of events for which nearly maximal spins can be ruled out, we see that this numbers is close to 90\% if only small primary spin systems are used. However, the curve is roughly flat. As we move to large $a_1^{max}$ we basically consider the whole distribution of spins, and obtain the same results of the left side of Fig.~\ref{Fig.SpinMin}.
It is worth stressing that the efficiency at excluding large spins is nearly always larger than for excluding small spins, the opposite only happening when the spins are in fact large. 
This is of course yet another way of saying that it's easier to measure large spins than small ones.

Given the relatively high fraction of events for which large spins can be excluded if the underlying population has random spins in the range $[0,1]$, it is thus not surprising that a similar conclusion could be drawn for \TheEvent{}.

In Fig.~\ref{Fig.SpinScatter} we explicitly show the 90\% CI (as errorbars around the median) for all our events, sorted by the true value of the primary spin (empty diamonds). As mentioned above, we see that even for large spins it is not uncommon that the posterior is centered around medium spins.

\begin{figure}[htb]
\includegraphics[width=0.95\columnwidth]{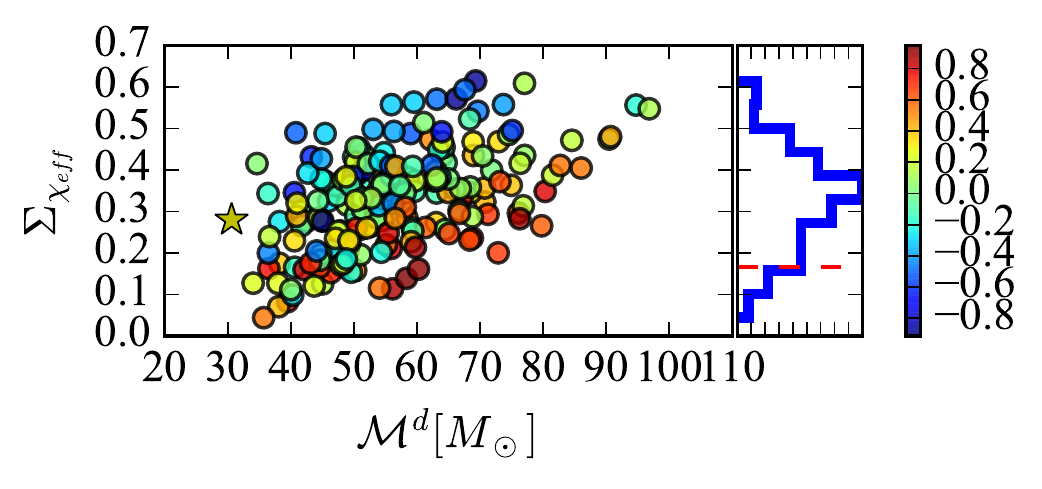}
\caption{The distribution of the 90\% CI uncertainty in the estimation of \chieff{} (y axis) against the true detector-frame chirp mass (x axis). The colorbar shows the true \chieff. A star reports the coordinates of \TheEvent{}. The dashed line on the histogram marks the abscissa of the $10^{th}$ percentile.}
\label{Fig.ChiHist}
\end{figure}

\begin{figure}[htb]
\includegraphics[width=0.95\columnwidth]{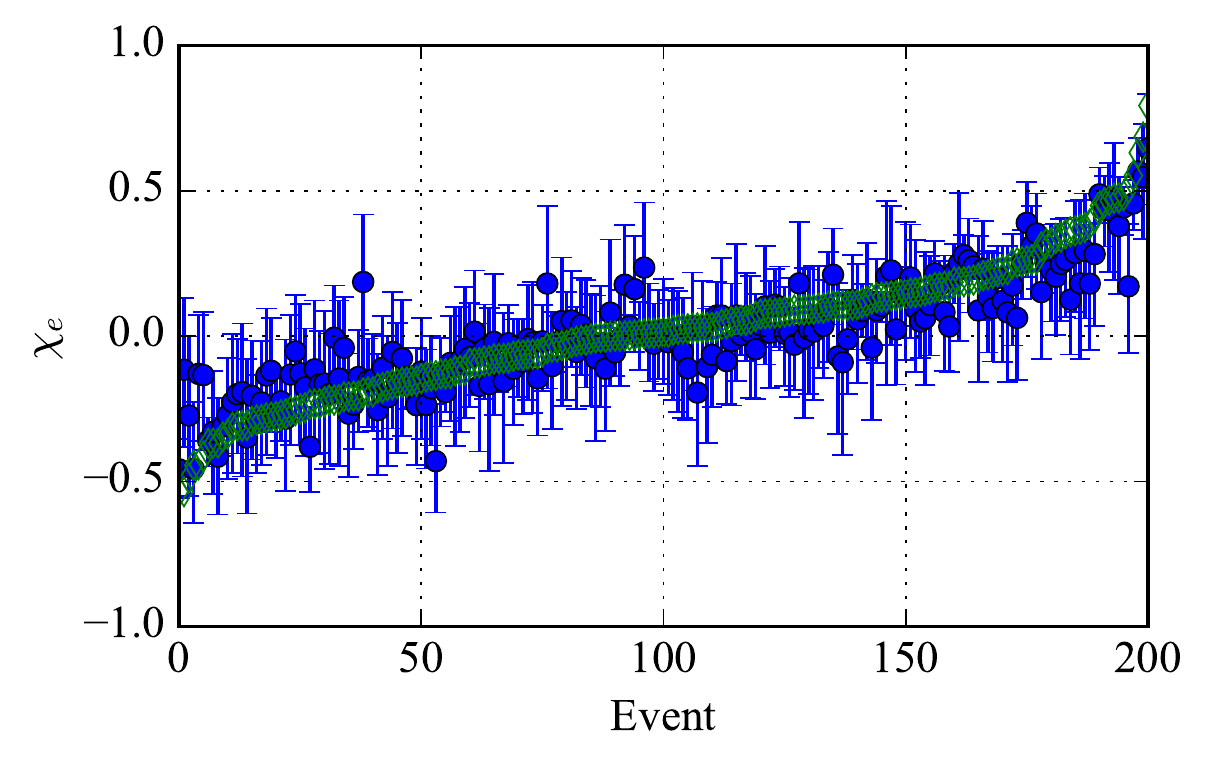}
\caption{For all events, median value of \chieff{} 90\% CI. The empty green diamonds indicate the true values.}
\label{Fig.ChiEffScatter}
\end{figure}

\begin{figure}[htb]
\includegraphics[width=0.95\columnwidth]{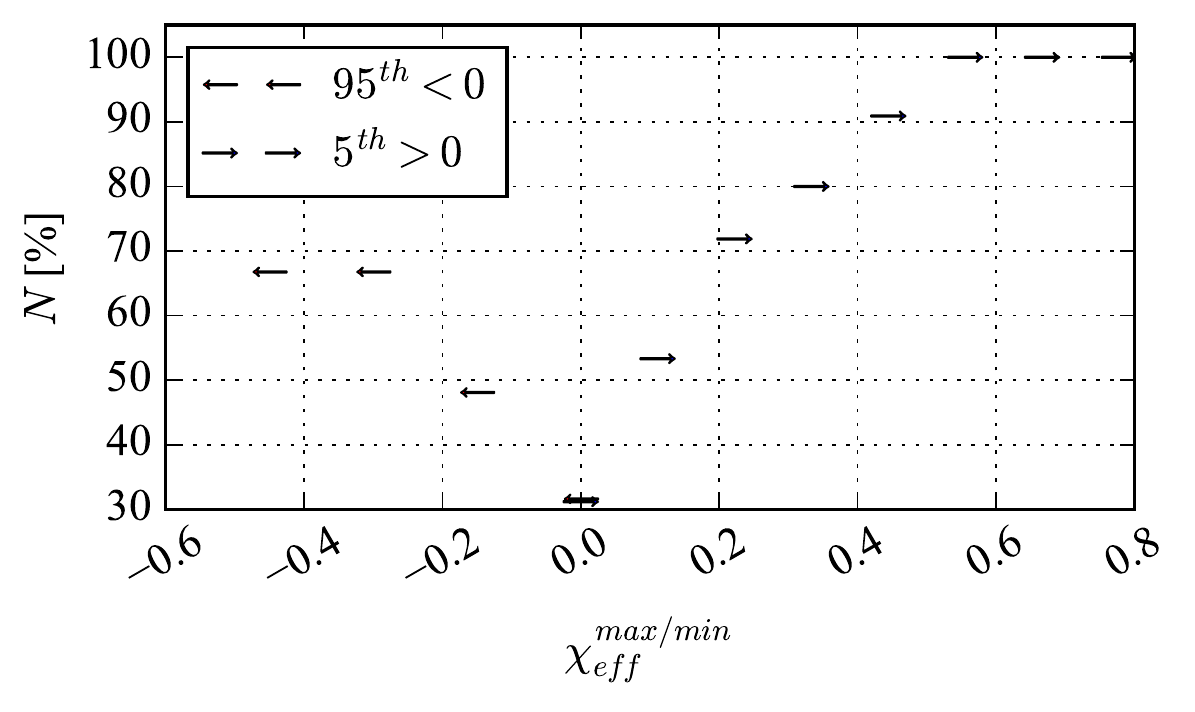}
\caption{The arrows pointing to the left report the fraction of events with true \chieff{} below that abscissa for which the $95^{th}$ percentile for \chieff{} is below zero. The arrows pointing to the right report the fraction of events with true \chieff{} above that abscissa for which the $5^{th}$ percentile for \chieff{} is above zero.}
\label{Fig.ChiEffMinMax}
\end{figure}

Let us now look at the estimation of the effective total spin along the orbital angular momentum. This is a quantity which was referred to as \chieff{} in \cite{GW150914-PARAMESTIM,O1-BBH}. Motivations for the use of this parameterization can be found elsewhere~\cite{PhysRevD.88.064007,PhysRevD.64.124013,PhysRevD.84.084037,PhysRevD.78.044021,PhysRevLett.106.241101,PhysRevD.82.064016}. Here we stress that being able to measure the sign of \chieff{} with high confidence could help favor some formation models for compact binaries~\cite{2016arXiv160905916R}. In fact, the main claim that could be made about the spins of \Xmas{} is that \chieff{} was positive and non-zero~\cite{O1-BBH,GW151226-DETECTION}. We find that \chieff{} is estimated better than either component spins. A similar conclusion was reached by \cite{PhysRevD.93.084042} for aligned-spin BBHs. 
In Fig.~\ref{Fig.ChiHist} we show the distribution of the 90\% CI for \chieff{} against the detector frame chirp mass. The colorbar reports the true \chieff. We see that the uncertainties are typically much smaller than what obtained while estimating the component spins (Figs.~\ref{Fig.SigmaA1_MC} and \ref{Fig.SigmaA2_MC}). This is not surprising, since it is the total spins, and in particular its projection along the orbital angular momentum, that affects the waveform length in both time and frequency domain. In particular, 10\% of events will have 90\% CI uncertainties below 0.17, with the typical event having uncertainties of \si0.35. For comparison, \TheEvent{} had a 90\% CI of 0.28~\cite{O1-BBH}.
In Fig.~\ref{Fig.ChiEffScatter} we show the median estimates for \chieff{} with the 90\% CI for all simulated events, with the green diamonds reporting the true simulated values. 
The small uncertainties suggest one might learn from \chieff{} more rapidly than from the component spins. We have verified how often one can exclude negative (positive) values for \chieff{} if the underlying population has positive (negative) true values, Fig.~\ref{Fig.ChiEffMinMax}. The arrows pointing to the left report the fraction of events having \chieff{} below the corresponding abscissa for which the 95$^{th}$ percentile of the \chieff{} posterior is negative. We see that when the populations has \chieff{} below -0.3, \si70\% of events can be correctly identified as having negative \chieff. The leftmost point is not reliable since very few events in our population have \chieff{} below -0.4. We expect that if the population extended to more negative values, the efficiency would continue to go up. 
We see this happening when we perform the opposite exercise (arrows pointing to the right). For example, if the population has positive \chieff{} larger than +0.3, 80\% of the times negative \chieff{} can be excluded.
Naturally, the exact numerical values of the efficiency at measuring the sign of \chieff{} depends on the population we simulated. However it seems safe to say that it is a much easier measurement than that of the individual spins.

We end this section with a quick discussion of tilt angles, i.e. the angle
between the spins and the orbital angular momentum. We will focus on the
primary object since, as for the spin magnitude, the tilt angle of the
secondary object will typically be unmeasurable.
The tilt angles are among the key quantities we wish to measure in a BBH, since
they could directly be linked to the formation channel of
CBCs~\cite{2015arXiv150304307V,2000ApJ...541..319K,RodriguezSpins}.
Of course, they are not constant during the evolution of the waveform, since both the spins and the angular momentum precess around the total angular momentum. Similarly to what done in \cite{GW150914-PARAMESTIM,O1-BBH}, we will quote the values of tilts at a frequency of 20~Hz.

\begin{figure}[htb]
\includegraphics[width=0.95\columnwidth]{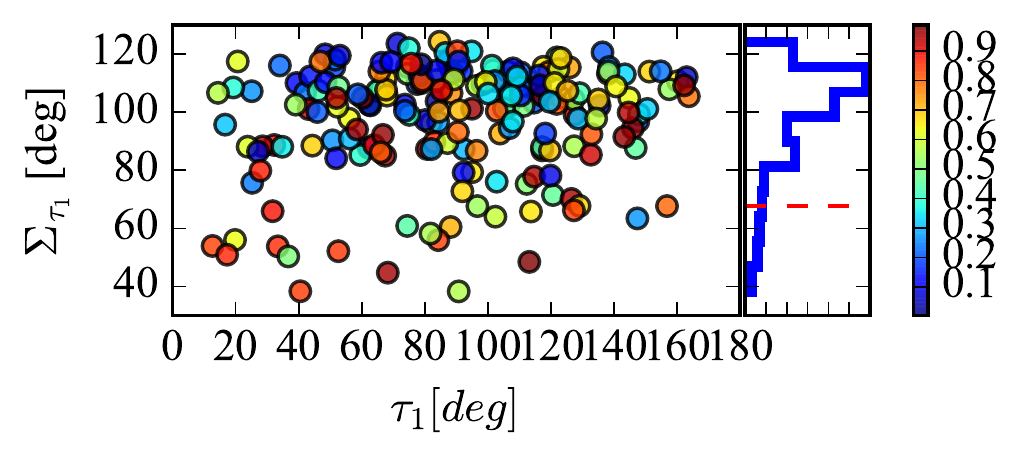}
\caption{The distribution of the 90\% CI uncertainty in the estimation of the
primary spin tilt angle (y axis, degs) against the true tilt angle (x axis,
degrees). The colorbar shows the magnitude of the primary spin. The dashed line on the histogram marks the abscissa of the $10^{th}$ percentile.}
\label{Fig.SigmaTau1_MC}
\end{figure}

In Fig.~\ref{Fig.SigmaTau1_MC} we report the 90\% CI for the tilt of the
primary, $\tau_1$ against its true value, both in degrees. The spin of the
primary is given in the color bar.  We see that for the typical event the
uncertainty will be very large: the distribution peaks at $\sim 110^\circ$
(histogram on the right panel). Only for $\sim 6\%$ of the systems will the
uncertainty be smaller than $60^\circ$. Once again, \TheEvent{} (for which we
don't show a star since the medians for the tilt angles were not made public)
fits perfectly in this scenario, since for none of the spins it was possible to
estimate the orientation~\cite{GW150914-PARAMESTIM}.

From Fig.~\ref{Fig.SigmaTau1_MC} we see that large spins are typically required
to have a chance of estimating the tilt angle. The other factor that plays a
large role in the capability of measure spins parameters is the orientation of
the orbital plane, which we discuss in the next section.

\subsection{Distance and sky location}\label{Sec.DistSky}

We end the analysis of the uncertainties of a population of BBH events with the luminosity distance and sky location. 
Precise estimation of distance and sky position will play a role in some of the proposed methods to calculate cosmological parameters with gravitational waves and to pinpoint the host galaxy of CBC sources~\cite{2012PhRvD..86d3011D,2010CQGra..27u5006S,2016ApJ...829L..15S}.

\begin{figure}[htb]
\includegraphics[width=0.95\columnwidth]{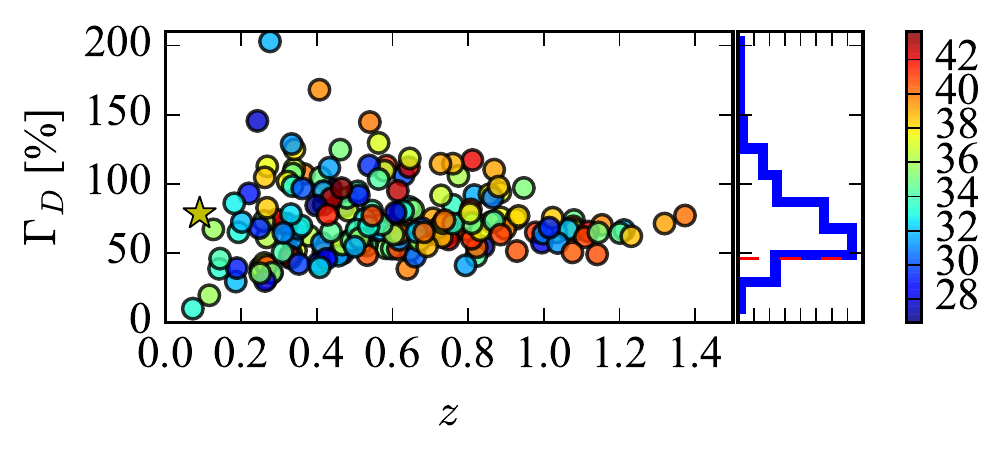}
\caption{The distribution of the 90\% CI relative uncertainty (in percent over the true value) in the estimation of the luminosity distance (y axis) against the true redshift (x axis). The colorbar is the source-frame chirp mass of the sources, in~\msun. A star reports the coordinates of \TheEvent{}. The dashed line on the histogram marks the abscissa of the $10^{th}$ percentile.}
\label{Fig.SigmaDist_z}
\end{figure}

In Fig.~\ref{Fig.SigmaDist_z} we show the relative 90\% CI uncertainty against the true redshift, the color reports the true source frame chirp mass.
We see that uncertainties have scatter for low distances, then converge toward values of around 50\%.
A rough Fisher matrix based approach would suggest that the relative errors should only depend on the SNR~\cite{1994PhRvD..49.2658C,2005PhRvD..71h4008A}. Since for large redshifts most events will have similar SNR (corresponding to the threshold 
 value we used to consider an event ``detected"), that explains why the points converge to a similar value.

We find that the uncertainties peak at \si50\%, slightly below what was found for GW150914.
 
We should stress that we are only reporting statistical uncertainties in the luminosity distance. 
As LIGO and Virgo start to detect sources at non-negligible redshifts weak lensing could affect distance measurement.
This potential systematic effect has already been investigated in the contest of third-generation gravitational waves detectors, such as the Einstein Telescope~\cite{2010CQGra..27s4002P} or the Cosmic Explorer~\cite{2016arXiv160708697A,2014arXiv1410.0612D}. Following \cite{2010CQGra..27u5006S}, we can assume that weak lensing could introduce a systematic of \si5\% on the the luminosity distance measurement for sources at z=1, and smaller for sources at smaller redshifts. For all the sources in our study this potential systematic effect would thus be much smaller than the statistical uncertainty.

While unlikely, it is not impossible that BBH will in fact emit energy in the electromagnetic band, or neutrinos, as some mechanisms have been proposed~\cite{2016ApJ...819L..21L,2016ApJ...822L...9M} after the discovery of \TheEvent{} and the potential EM sub-threshold trigger found by the Fermi mission~\cite{2016arXiv160203920C}. 
Furthermore, it could be possible to use the position of detected events to study the large-scale structure of the Universe~\cite{2012PhRvD..86h3512J,2016PhRvD..94b4013N} and to look for the host galaxy and calculate the cosmological parameters~\cite{2012PhRvD..86d3011D}.

\begin{figure}[htb]
\includegraphics[width=0.95\columnwidth]{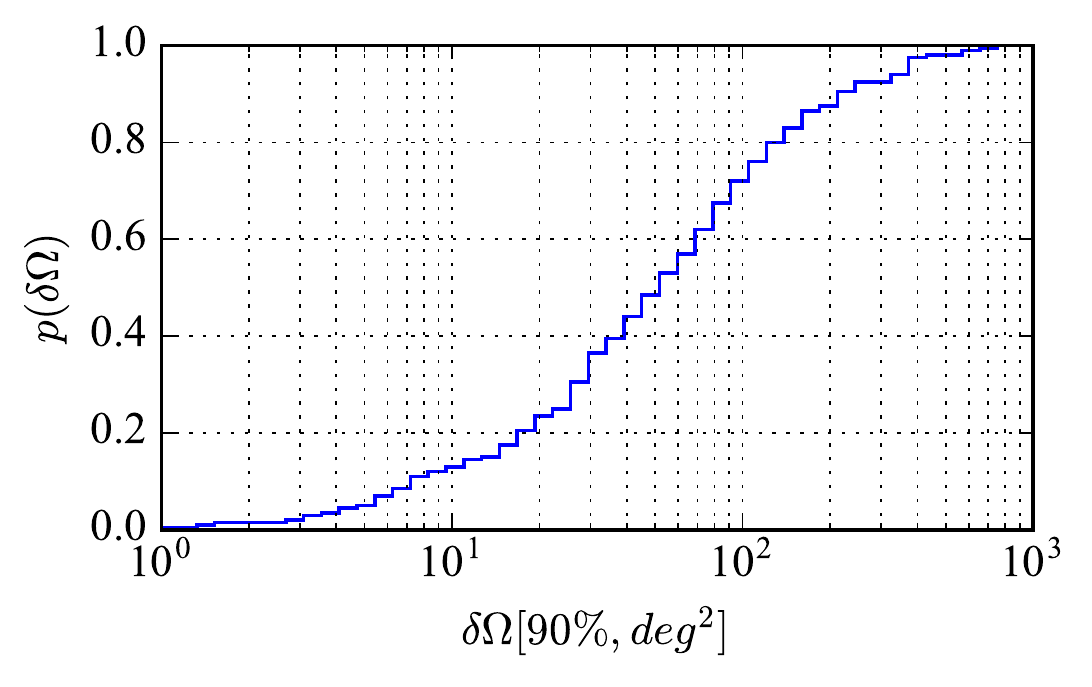}
\caption{Cumulative distribution for the 90\% CI sky localization area for BBH detected by the HLV network.}
\label{Fig.Sky}
\end{figure}

In Fig.~\ref{Fig.Sky} we show cumulative distributions for the 90\% credible interval for the sky position, in square degrees. 
In our runs we have not included marginalization over instrumental calibration uncertainties, which have the potential to increase the sky uncertainties~\cite{O1-BBH,GW150914-PARAMESTIM} or to bias it, if not accounted for~\cite{2012PhRvD..85f4034V}.
We have implicitly assumed that by the time the advanced detectors reach design sensitivity, calibration uncertainties, which are now at a \si5\% level~\cite{GW150914-CALIBRATION,O1-BBH}, will be better understood.

Our results are comparable with \cite{VeitchMandel:2012}, which focused on binary neutron stars. The main difference is that the uncertainties we obtain for BBH are larger than what they obtained for binary neutron stars, in spite of the fact that we quote 90\% CI, while they used 95\% CI.

For example, for the HLV network (actually HHLV in \cite{VeitchMandel:2012} since they considered the two detectors that were at the Hanford site, one of which will be relocated to India) we obtain a median uncertainty of 50~deg$^2$, while \cite{VeitchMandel:2012} obtains \si$30$~deg$^2$.

This is, of course, due to the fact that BBH signals have a smaller effective bandwidth~\cite{2011CQGra..28j5021F}, and hence harder to localize than longer binary neutron star sources~\cite{2011PhRvD..84j4020V}.

\sv{Finally, it is worth mentioning that the sky maps shared with partner astronomers for prompt follow-up are currently produced by a low-latency algorithm (BAYESTAR,~\cite{2016PhRvD..93b4013S}), while \linf{} sky maps \jv{which include a more detailed model of the source and instrument calibration} follow with an higher latency. It has been shown that the low and medium latency maps are in very good agreement for a network of two instruments, while the agreement is lower for a three-instrument network \jv{because \linf{} is able to use data from all three detectors regardless of the presence of a trigger}~\cite{2014ApJ...795..105S}. This discrepancy is currently being addressed in preparation for Advanced Virgo's first observing run~\cite{LeoBS} (see Section X of~\cite{2016PhRvD..93b4013S}). Ref.~\cite{2014ApJ...795..105S} deals with binary neutron stars, but the situation should be similar for BBH, unless significant spin precession is present. In that case \linf{} should provide a more accurate skymap, since the low-latency algorithm \jv{is based on the output of search pipelines which} currently neglect precession.}

\section{Trends}

In the previous section we have focused on an astrophysical population of events and obtained distributions for the expected uncertainties of the sources' parameters.
We now want to show how the estimation of the spin parameters depends on the intrinsic parameters of the source (i.e. mass and spin) as well as on its orientation.

\subsection{Dependence on orientation}

It is commonly assumed that in the limit of spins aligned with the orbital angular momentum, the spin parameters are strongly degenerate with the mass ratio for small masses~\cite{2013PhRvD..87b4035B} and hence hard to measure. Mathematically, this happens because the leading order spin term in the waveform inspiral phase depends on a combination of mass ratio and (aligned) spins.
At the same time, when misaligned spins are present, spin-spin and spin-orbit interactions will make the orbital plane precess, which gives the signal amplitude and phase modulation~\cite{PhysRevD.49.6274}. 
One would thus think that precessing spins are easier to measure. Since the amount of precession visible at Earth is also a function of the inclination angle~\cite{PhysRevD.49.6274,PhysRevLett.112.251101,GW150914-PARAMESTIM}, the best case scenario should be when precession is present and the system is observed from ``edge-on" (orbital angular momentum forming an angle of $\pi/2$ with the line of sight).
In \cite{PhysRevLett.112.251101} it was shown, for one particular low-mass BBH system, how uncertainties in the measurement of spin do indeed reach a minimum for inclination angles close to $\pi/2$.

However, as \cite{2013PhRvD..87b4035B} underlines, there is no reason the known degeneracies of the inspiral phase should hold true when the merger and ringdown parts of the waveform are measurable.

In this section we investigate how the characterization of heavy BBH sources depends on the orientation of the system and their spins.

We consider several systems with different values of masses, spin orientation and SNR, and analyze them for different values of inclination angle (to be exact, what we have varied is the angle between the total angular momentum and the line of sight, $\theta_{JN}$~\cite{2014PhRvD..90b4018F,2015PhRvD..91d2003V}). 
The parameters of the sources we used are reported in Tab.~\ref{Tab.Rotate}. We consider mass ratios from 1 to 1:2.5, and mostly focus on large spins. Tilt angles are typically chosen large to be sure there are precessional effects to be seen in the first place.

\begin{figure}[htb]
\includegraphics[width=0.99\columnwidth,height=6cm]{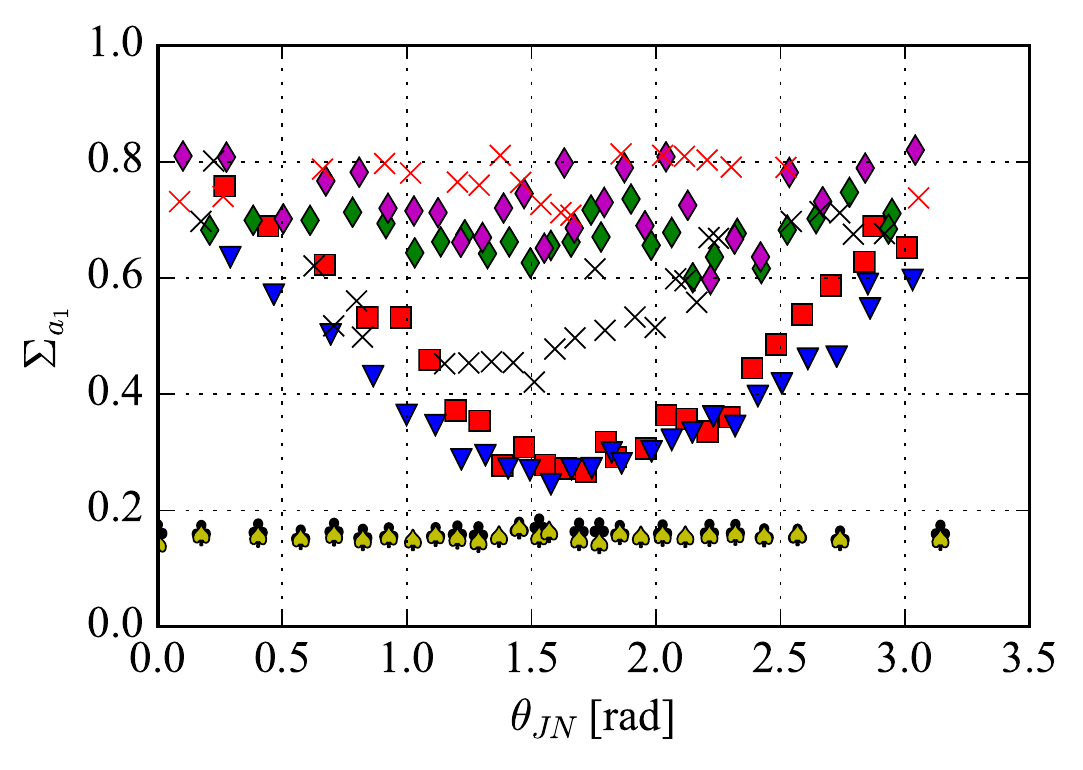}
\caption{90\% CI uncertainty in the primary spin magnitude against \tjn (rads) for various BBH sources. See Table~\ref{Tab.Rotate} for the parameters of the sources.}
\label{Fig.RotateA1}
\end{figure}

We stress that every time a system is rotated, its distance is varied to keep the same SNR. The variations we see are thus not due to variations in the loudness of the source, but only on the extra complexity of the signal when not face-on. In Fig.~\ref{Fig.RotateA1} we report the 90\% CI uncertainty for the primary spin against \tjn.

We see that the effect strongly depends on the mass ratio of the systems. For equal-mass sources (diamonds) we don't see any strong variation on the ability of measure the spin magnitude. This is compatible with the fact that spin induced modulation effects are minimal for equal-mass systems~\cite{2015PhRvD..91b4043S}. As the mass ratio increases, so does the effect of the inclination angle. For the source with mass ratio 1:1.5 (crosses) we start to observe a reduction of the uncertainties for large inclination angles, unless the spins are small.

The improvement is even more pronounced for the sources with mass ratios of $2$ (squares) and $2.5$ (triangles). For these sources, as expected, uncertainties reach their minimum for angles close to $\pi/2$. Furthermore, we see that the ratio between uncertainties in the best and worst case scenario can be over a factor of two.
Although in this paper we don't deal with neutron star - black hole binaries, the ratio would be even larger for those sources given their larger mass ratio.
We stress that by using \WF{} for large inclination angles we are in fact working on a corner of the parameter space where that approximant might not be highly reliable~\cite{PuerrerPrep}. The fact that the curves we obtain look similar to those reported in \cite{PhysRevLett.112.251101} using a different approximant (SpinTaylorT4~\cite{2003PhRvD..67j4025B,2006PhRvD..74b9904B}) and lower masses, reassures us that the results we find in this section are at least a good indication of the trends one can expect. Of course, a similar study should be repeated as soon as fast double-spinning IMR waveforms become available. Potential systematics against numerical relativity waveforms should also be quantified.

We now want to verify if aligned spins are harder to measure even for heavy BBH, for which merger and ringdown are in band.
In Fig.~\ref{Fig.RotateA1} we show results for two spin-aligned BBH, with mass ratio of 1 (black club suits) and 2 (yellow spade suits). In both cases, the spins are 0.9 (see Table~\ref{Tab.Rotate}). 
We stress that while the simulated BBH had aligned spins, the parameter estimation algorithm did \emph{not} make this assumption, i.e. we explored the full precessing parameter space.
We see that the uncertainties in this case are considerably smaller than all other precessing systems we considered, at around 0.2.

As mentioned above, it has been stressed elsewhere~\cite{2013PhRvD..87b4035B}  that one should not a-priori expect the same correlations found in inspiral-dominated (i.e. low mass) systems to hold true for heavy BBH. This is also consistent with the fact that for large and aligned spins the length of a waveform is increased~\cite{2006PhRvD..74d1501C}. While this effect would be degenerate with the total mass if only the inspiral phase were in band, the presence of a measurable merger and ringdown breaks that degeneracy, improving the measurability of spin parameters.

We have verified that these results hold true if a) the template used for parameter estimation only consider aligned-spin and b) the simulated signals are not exactly aligned but have tilt angles of \si3-5 degrees.

We thus find that uncertainties for spin-aligned heavy BBH can be much smaller than for precessing systems of similar masses if significant spin is present. Our finding that spins nearly aligned with the orbital angular momentum can be estimated very well when the merger and ringdown are in band is compatible with what found in \cite{PhysRevD.93.084042}, which used a different waveform family.

\begin{table}[htb]
\centering
\begin{tabular}{|c|c|c|c|c|c|} \hline
  & $a_1$,$a_2$ & $m_1$,$m_2$ & SNR & $\cos\tau_1$, $\cos\tau_2$ &marker\\
  \hline
q1a &   0.9,0.9      &      35,35      &  17    &     0.5,0.5  &\textcolor{OliveGreen}{$\blacklozenge$} \\  \hline
q1b &  0.9,0.2      &      35,35      &  25    &     0.8,0.5  &\textcolor{Fuchsia}{$\blacklozenge$} \\  \hline
q1d5 &   0.9,0.9      &      45,30      &  17    &      0.5,0.5& \textcolor{Black}{$\times$}    \\  \hline
q1d5ss &   0.4,0      &      45,30      &  17    &      0.5,0.5   &  \textcolor{Red}{$\times$} \\  \hline
q2a &      0.9,0.9      &      70,35      &  17    &      0.5,0.5    & \textcolor{Red}{$\blacksquare$} \\  \hline
q2d5 &      0.9,0.9      &      75,30      &  17    &      0.5,0.5   & \textcolor{Blue}{$\blacktriangledown$}  \\  \hline
q1ali &      0.9,0.9      &      35,35      &  17    &      1,1   & \textcolor{Black}{$\clubsuit$}  \\  \hline
q2ali &      0.9,0.9      &      70,35      &  17    &      1,1   & \textcolor{YellowOrange}{$\spadesuit$}  \\  \hline
\end{tabular}
\caption{Intrinsic parameters and network SNRs for the systems of Fig.~\ref{Fig.RotateA1}. Masses are in \msun.}
\label{Tab.Rotate}
\end{table}

\subsection{Dependence on mass}\label{Sec.DepOnMass}

The results of the previous section have shown how characterization of heavy BBH might have properties that were not previously thoroughly discussed or investigated.
In this section we want to investigate another common assumption, that heavier CBCs are harder to characterize, being shorter in both the time and the frequency domain.

We consider two precessing systems with fixed mass ratios of 1 (green diamonds) and 2 (red squares) and a spin-aligned system with mass ratio of 1 (black club suits). Their parameters are given in Table~\ref{Tab.Mass}.

\begin{table}[htb]
\centering
\begin{tabular}{|c|c|c|c|c|c|} \hline
  & $a_1$,$a_2$ & $q$ & SNR & $\cos\tau_1$, $\cos\tau_2$ &marker\\
  \hline
q1 &   0.9,0.9      &      1     &  17    &     0.5,0.5  &\textcolor{OliveGreen}{$\blacklozenge$} \\  \hline
q2 &      0.9,0.9      &      2      &  17    &      0.5,0.5    & \textcolor{Red}{$\blacksquare$} \\  \hline
q1ali &      0.9,0.9      &      1      &  17    &      1,1   & \textcolor{Black}{$\clubsuit$}  \\  \hline
\end{tabular}
\caption{Intrinsic parameters and network SNRs for the systems of Fig.~\ref{Fig.IncreasMassA1}. Masses are in \msun{} and \tjn{} is $45$~degs for all systems.}
\label{Tab.Mass}
\end{table}

These systems were simulated with increasingly large \emph{detector frame} total mass. Every time the total mass if varied, the distance to the source is also changed to yield the same network SNR for all masses. It must be recalled than when spin-induced orbital precession is present some spin parameters become time, and hence frequency, dependent. Throughout this work we have defined spin parameters, such as the tilt angles, at 20~Hz. 
However, in this section we make a different choice. To ensure that spins are defined at fixed number of cycles before merger, we define spins at a different reference frequency for each value of mass. To be precise, for each $M_{tot}$ the spins are defined at a reference frequency, such that $M_{tot} f_{ref}= const$.

\begin{figure}[htb]
\includegraphics[width=0.99\columnwidth,height=6cm]{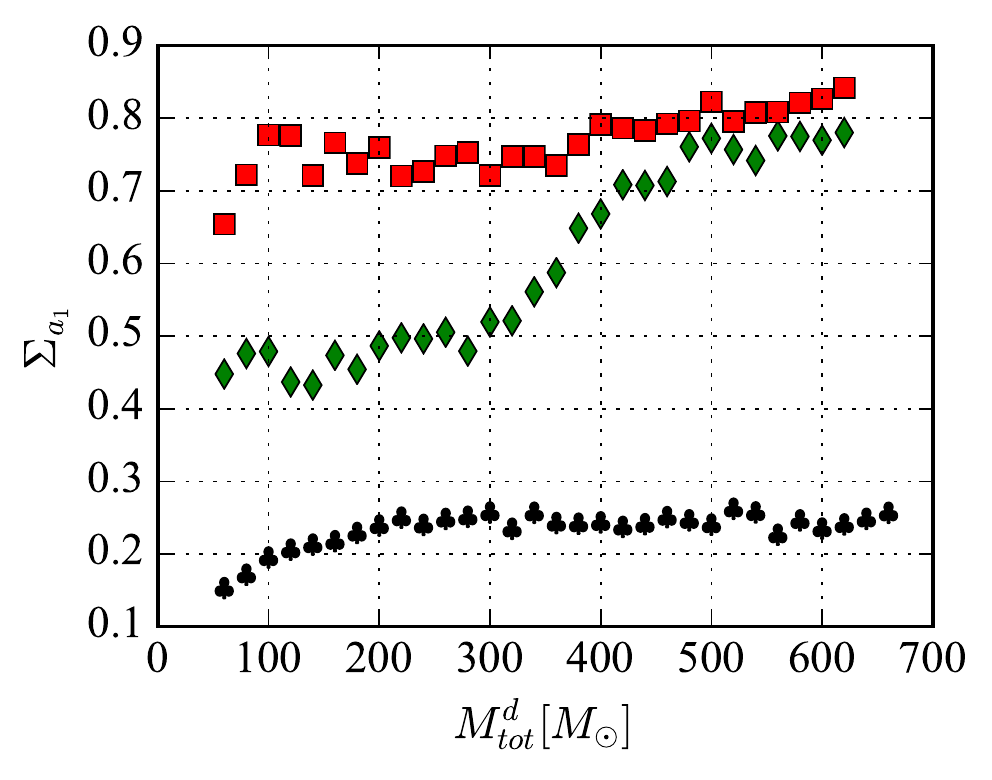}
\caption{90\% CI uncertainty in the primary spin magnitude against the detector frame total mass. See Table~\ref{Tab.Mass} for the parameters.}
\label{Fig.IncreasMassA1}
\end{figure}

We first look at the estimation of the magnitude of the primary spin. In Fig.~\ref{Fig.IncreasMassA1} we show the 90\% CI for the primary spin magnitude versus the redshifted total mass.
We see how, while the overall trend is an increase of the uncertainties with the total mass, the amount of variation depends on the mass ratio. 
The precessing equal-mass system (red squares) shows the smallest variation, with uncertainties which are significantly large already at small masses. On the other hand, the system with mass ratio of 2 has mild uncertainties at $M=60$~\msun which increase by a factor of 2 as the total mass increases to $M=600$~\msun.
Remarkably, the uncertainties for the spin-aligned system (club suits) stays much smaller than the precessing spins systems in the whole mass range.

\begin{figure}[htb]
\includegraphics[width=0.99\columnwidth]{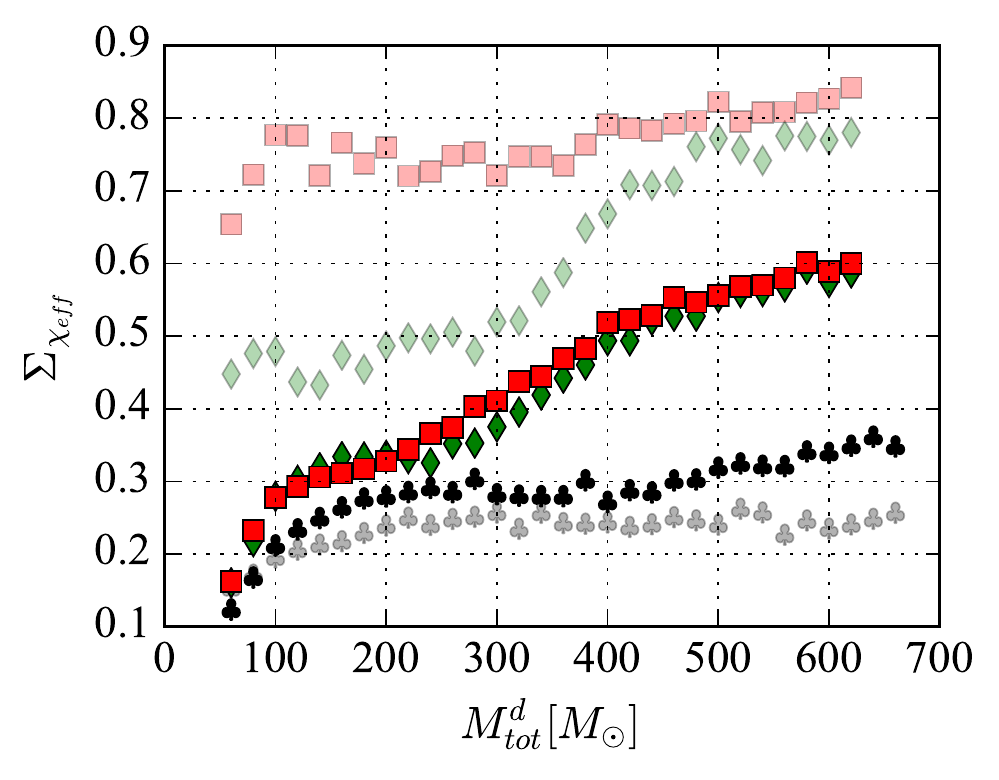}
\caption{90\% CI uncertainty in the effective spin along the orbital angular momentum against the detector frame total mass. The uncertainties in primary spin magnitude are also shown (semi-transparent symbols) for reference. See Table~\ref{Tab.Mass} for the parameters.}
\label{Fig.IncreasMassChiEff}
\end{figure}

Next, we report the uncertainties on the measurement on the effective spin along the orbital angular momentum. As we have seen above, the effective spin parameter can generally be estimated more precisely than either component spin. We find this is the case for all values of masses we consider, at least for the precessing systems, Fig.~\ref{Fig.IncreasMassChiEff}. For the spin-aligned system we see that the uncertainty in $\chi_{eff}$ is similar to the uncertainty in $a_1$, which is not surprising since the whole spin is along the orbital angular momentum, and hence contributes to the effective spin.

\begin{figure}[htb]
\includegraphics[width=0.99\columnwidth]{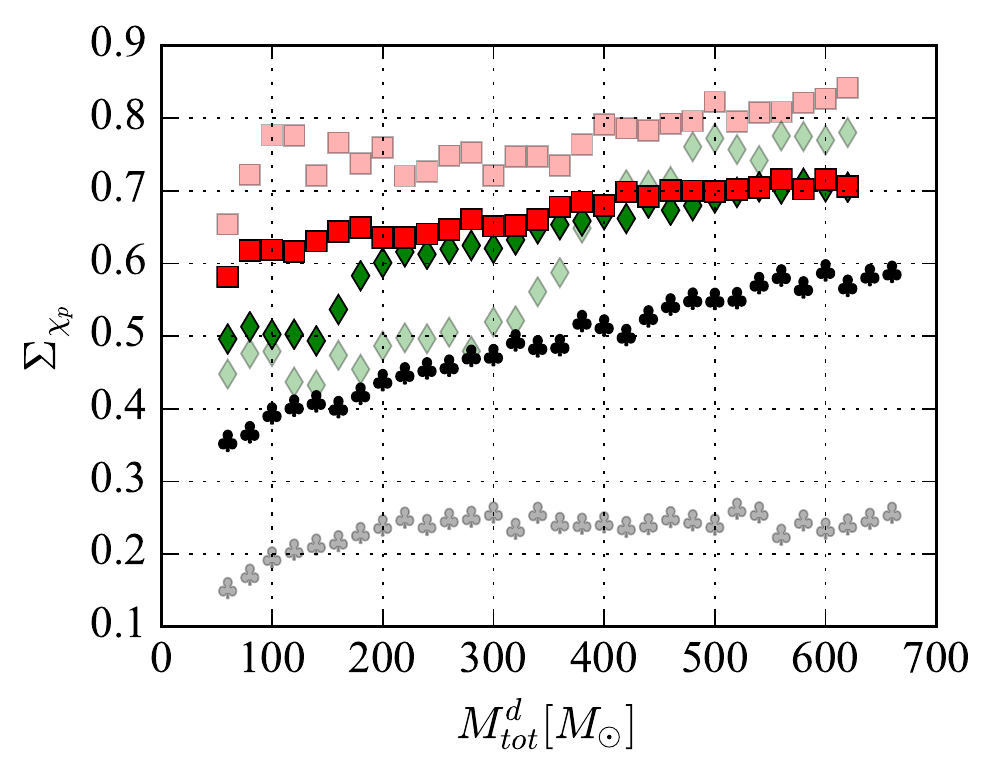}
\caption{90\% CI uncertainty in the effective spin along the plane of the orbit against the detector frame total mass. The uncertainties in primary spin magnitude are also shown (semi-transparent symbols) for reference. See Table~\ref{Tab.Mass} for the parameters.}
\label{Fig.IncreasMassChiP}
\end{figure}

We notice that the uncertainty in the estimation of $\chi_{eff}$ is similar among the two precessing systems, while we had observed large differences in the measurement of the primary spin magnitude, Fig.~\ref{Fig.IncreasMassA1}.
This is due to the fact that the measurements of the component spins magnitude are also affected by the correlation of spin magnitude with spin orientation, which depends on how much precession is ``visible".
We thus look at the estimation of the effective precessing spin, $\chi_p$, i.e. a mass-weighted combination of the total spin component in the plane of the orbit. As for $\chi_{eff}$, motivations for the use of this parameterization has been discussed elsewhere~\cite{PhysRevD.88.064007,PhysRevD.64.124013,PhysRevD.84.084037,PhysRevD.78.044021,PhysRevLett.106.241101,PhysRevD.82.064016}. This is shown in Fig.~\ref{Fig.IncreasMassChiP}. 

We see that, especially for low masses, the $q=2$ system has smaller uncertainties for $\chi_p$ than the precessing equal-mass source. This is due to the fact that, as mentioned above, precession effects are more visible when there exit a mass asymmetry.  We have verified that the small jump in uncertainty for $M\sim 150$~\msun{} happens as the peak of the first precession cycle (i.e. the one at lower frequency) starts going out of band, due to the increasing total mass. 
We notice that for $\chi_p$ too, the spin-aligned system have smaller uncertainties. However that does not happen because they do better than the precessing systems. Quite the contrary, we find that the posterior for $\chi_p$ is centered at \si0.4 for most spin-aligned runs (the injected value is 0.0), not far from where the prior is centered. We notice however that the posterior for the spin-aligned runs is slightly narrower than the prior.

\section{Characterization of mass and spins distribution}

Although measuring the spins of \emph{single} objects will be hard, we stress that it will be possible to learn something about the underlying population combining information from several sources. 
For example, in Fig.~\ref{Fig.SpinMin} we have seen how one can very often discard large values of spins, if the true distribution has smaller spin values. In that case, one can imagine how as more detections are made large spins become less and less supported by the data.

In this section we want to use a very simple toy model to show how inference about the mass and spin distributions can be done.
Let us consider a set of 105 BBH with masses uniformly distributed  in the range  $[30,50]$~\msun{} and spins uniformly distributed in $a\in[0.7,0.98]$.
Under the hypothesis that mass and spin distributions are flat, with unknown boundaries, can the extrema be estimated? If yes, how many detections are needed?

Let us start by estimating the boundaries of the component masses distribution.
We will call $m_{min}$ and $m_{max}$ the minimum and maximum of the astrophysical distribution, and $\mathcal{H}$ the model that the distribution is flat. If N detections are made, symbolized by their data streams $\vec{d}$, then using Bayes' theorem one can write:

\begin{widetext}
\begin{equation}\label{Eq.MassPostDistro}
p(m_{min},m_{max} | \vec{d}, \mathcal{H})  \propto p(m_{min}, m_{max} |  \mathcal{H}) p( \vec{d} | m_{min}, m_{max} , \mathcal{H})   = p(m_{min}, m_{max} |  \mathcal{H}) \prod_{i=1..N}{ p( {d}^i | m_{min}, m_{max},\mathcal{H} )},
\end{equation}
\end{widetext}
where $d^i$ is the data stream of the $i-th$ signal.

Each term in the product is just the usual evidence of the data, but restricted to mass values between the min and max being considered.
This can be implemented trivially in the parameter estimation algorithm we used by restricting the prior range of the component masses~\cite{2015PhRvD..91d2003V}. In practice, to avoid wasting computational resources and since the original priors are flat, we just used importance sampling~\cite{Bishop:2006:PRM:1162264}.
The other term in the RHS is the prior distribution for the minimum and maximum, which we can take as flat.

\begin{figure}[htb]
\includegraphics[width=0.95\columnwidth]{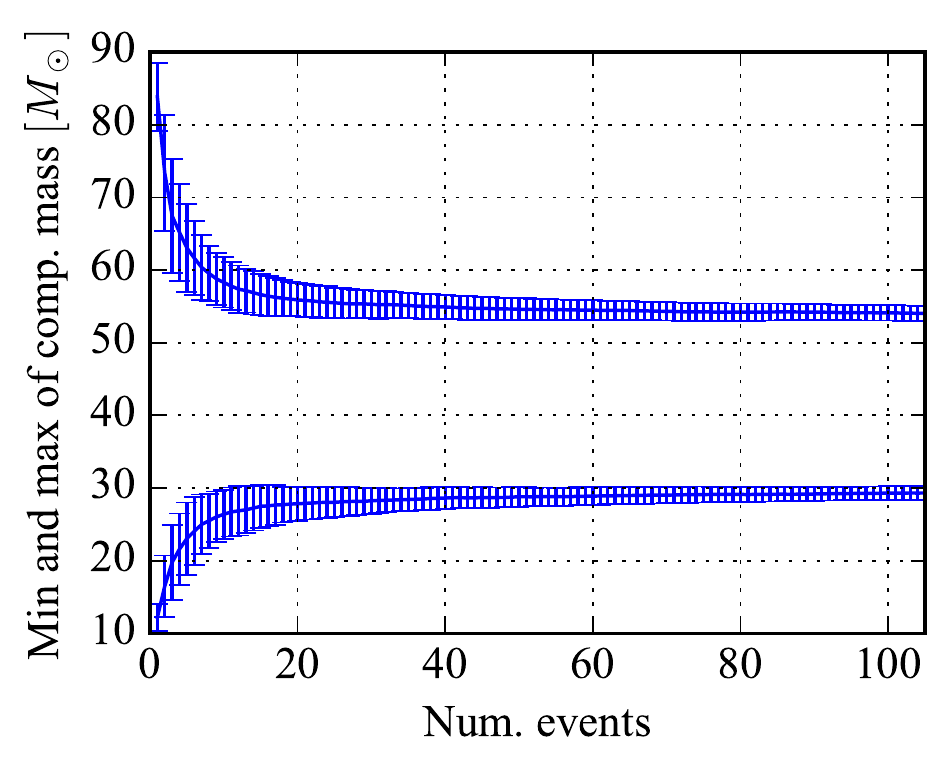}
\caption{The estimation of the minimum and maximum value of the component mass for a population of BBH as more events are detected. The errorbars are obtained bootstrapping 100 sets of N events for each value of N, in the x axis. The true distribution is flat between 30~\msun{} and 50~\msun{} for both masses. After 20 events, we can confidently measure the edges of the underlying mass distribution. }\label{Fig.MassesMinMax}
\end{figure}

In Fig~\ref{Fig.MassesMinMax} we show how the estimation of the minimum and maximum range for the source-frame component masses evolves as more events are detected. The x axis reports the number of events used, and the y axis the estimated values of the maximum (upper curve) and minimum (bottom curve). To calculate errorbars, for each choice of the number of events, N, we generated 100 random sets of N events with bootstrapping and calculated mean and standard deviation of the edges of the 90\% credible interval.

\begin{figure}[htb]
\includegraphics[width=0.95\columnwidth]{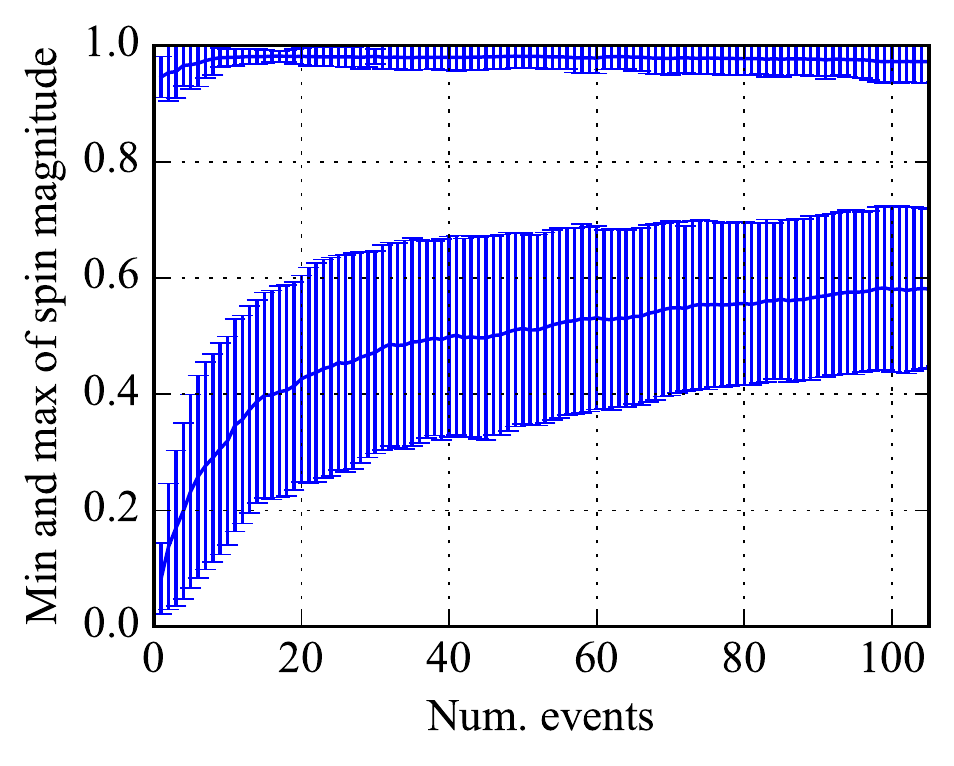}
\caption{The estimation of the minimum and maximum value of the spin magnitude for a population of BBH as more events are detected. The errorbars are obtained bootstrapping 100 sets of N events for each value of N, in the x axis. The true distribution is flat between 0.7 and 0.98. After 20 events, we can exclude non-spinning BBH in the underlying astrophysical population.}\label{Fig.SpinsMinMax}
\end{figure}

The same exercise can be done for the spin magnitude. Using the same expression we derived for the masses, one obtains the joint distribution for $a_{min}$ and $a_{max}$. 
In Fig.~\ref{Fig.SpinsMinMax} we show the evolution of the estimation for $a_{min}$ and $a_{max}$ as function of detected events.

We see that the error bars are much larger than for the masses, which is simply a consequence of the fact that spins are harder to measure than masses. After 10-20 events, non-spinning BBH are excluded, and after a few tens the data points to a minimum spin at around 0.6, with standard deviations of \si0.1.

\sv{The results of this section should be seen as the simple application of a toy model, and are only meant to give the reader an idea of what can be done when several sources are available. 
Here we list \vivien{three} main caveats.
The number of sources that are needed to e.g. exclude negligible spins are of course dependent on our choice \jv{of population, of which we consider one possibility}. For example, if the true population had spins down to e.g. 0.4 rather than 0.7 then more sources would be needed.
Furthermore the true astrophysical distribution spin and mass would not have sharp boundaries and its shape would not be known to start with. Measuring the edges of a flat (top-hat) distribution lead to better results than estimating the parameters of more realistic distributions (e.g. gaussian, power law)~\footnote{We thank Will Farr for having clarified this point.}. An example of a more elaborate treatment in the context of population modeling can be found in \cite{2016arXiv160808223M}, which appeared when this work had already started.}
\vivien{Finally, as mentioned in section~\ref{Sec.Events}, the \WF{} approximant may not be able to accurately compute the gravitational waveform for the few edge-on~\cite{VitaleDiff,2011CQGra..28l5023S}, low SNR, signals in the population.}

\section{Conclusions}\label{Sec.Conclusions}

In this paper we have considered an astrophysical distribution of heavy ($m_{1,2} \in [30,50]~\msun$) spinning BBH detected by a network of advanced LIGO and Virgo detectors.
Sources like these will be detected in high number in the next few years, and it is interesting to verify what kind of measurement one can expect for the masses and spins of black holes in these systems.

We find that source-frame component masses will be estimated with typical relative uncertainties of the order of \si40\%.  The exact size of the errors will depend, beside the signal-to-noise ratio, on the detector-frame masses, since those control the duration and amplitude of the signal.
There will thus be a coupling between source-frame mass estimation and source redshift. This correlation will be exacerbated in the next generation of gravitational wave detectors~\cite{Vitale3G}.
The source-frame chirp mass is estimated with similar precision.

The spin magnitude of either object in the binary will typically be estimated with large uncertainties. We found that for the primary (i.e. most massive) object in the system only 10\% of sources will yield a measurement with uncertainty below 0.7. For the secondary, below 0.85. We found that large spins typically can be estimated with smaller uncertainties, similarly to what happens for BH in X-ray binaries. The effective spin along the orbital angular momentum,~\chieff, can be measured better than either spins, with uncertainties for 10\% of sources below 0.17.

Considering only the BBH in our population with primary spin below 0.2, we saw that \si90\% of the times one can exclude that the BH was fast spinning (i.e. with spin above 0.95). This number goes down to roughly 80\% if a flat distribution of spin is used. Conversely, if only BBH with primary spin above 0.8 are used, 75\% of them will not support negligible spins (i.e. spin below 0.05). If the whole flat spin distribution is used, 55\% of the systems will exclude negligible spins. 
We have checked how well the sign of the effective spin can be measured, which could be used to prefer some formation models for CBCs. We have found that if one only considers BBH with \chieff$<-0.3$ (\chieff$>+0.3$) 70\% (80\%) of the times one can exclude positive (negative) \chieff.

The angle between spin and orbital angular momentum, which could also be used to probe the formation channels of CBC, will also be estimated quite poorly. For only 6\% of our BBH the 90\% CI for this angle is below 60\degs.

We have verified that the uncertainties of \TheEvent{} for both masses and spins are typical of events in the same mass range. We have shown how correlations can exist between the ability of measuring the spin parameters, for precessing systems, and the inclination of the orbit. However, these correlations are only clear if the mass ratio is not close to unity. For equal mass systems, precessing spins are hard to measure no matter of the orientation of the orbit.
We considered spin-aligned systems with mass ratio of 1 and 2 and spin magnitude of 0.9, and found that the spin magnitude can be measured extremely well, with 90\% CI of \si0.2. This is contrary to what traditionally expected for low mass CBCs, which are dominated by the inspiral phase, and show a strong degeneracy between spin and mass ratio.

We then investigated how the uncertainties on the spin magnitude depend on the detector frame total mass. We found that while uncertainties get larger overall for larger masses, the increase is much more significant when the mass ratio is not close to unity. For the system with mass ratio of 2 we considered, the uncertainty in the primary spin magnitude at $M_{tot}=60$~\msun{} is a factor of 2 smaller than at  $M_{tot}=600$~\msun.

Finally, we have verified what can be said about the mass and spins of the underlying distribution of BBH events. Considering a toy model where masses and uniform in the range $[30-50]$~\msun{} and spins uniform in the range $[0.7-0.98]$, we have shown how the boundaries can be measured assuming a top-hat distribution, with less than 100 detections.
A top-hat distribution is of course only a crude approximation, and more work will needed to assess the characterization of more realistic distributions.

\section{Acknowledgments}

The authors would like to thank T.~Dent, D.~Gerosa, V.~Kalogera, M.~P\"urrer, C.~Rodriguez, R.~O'Shaughnessy and the LSC-Virgo parameter estimation subgroup for useful discussion and comments. We also thank the Referee for the many useful comments.
SV and RL acknowledge the support of the National Science Foundation and the LIGO Laboratory. LIGO was constructed by the California Institute of Technology and Massachusetts Institute of Technology with funding from the National Science Foundation and operates under cooperative agreement PHY-0757058. JV was supported by STFC grant ST/K005014/1.
During this work RS has been supported by FAPESP grants 2012/14132-3, 2013/04538-5 and 2014/50727-7.
The authors would like to acknowledge the LIGO Data Grid clusters, without which the simulations could not have been performed. Specifically, we thank the Albert Einstein Institute in Hannover, supported by the Max-Planck-Gesellschaft, for use of the Atlas high-performance computing cluster. We are also grateful for computational resources provided by the Leonard E Parker Center for Gravitation, Cosmology and Astrophysics at University of Wisconsin-Milwaukee.

This is LIGO Document P1600292. 

\appendix
\section{A 5-detector network}\label{App.HLVIJ}

In this appendix we report results on a different BBH population, with intrinsic component masses flat in the range \HLVIJMASSRANGE{} (and $M_{tot}\leq 100$\msun) as detected by a 5-detector network  which  includes the two LIGO, Virgo, KAGRA and LIGO India (henceforth HVLIJ). The main goal of this section is to show that, if an astrophysical distribution of BBH of roughly similar masses is considered, the actual configuration of the network do not matter, in first approximation, for the measurement of the intrinsic parameters.
We will in fact see that the uncertainties we obtain with the HLVIJ network are similar, for masses and spins, to what we reported in the main body for the smaller HLV network.

\begin{figure}[htb]
\includegraphics[width=0.95\columnwidth]{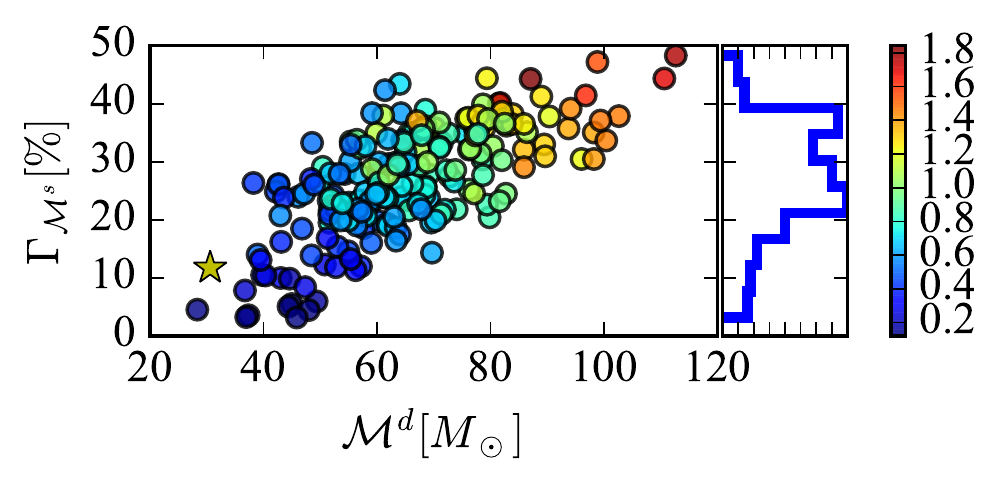}
\caption{For an HLVIJ network, the distribution of the 90\% CI relative uncertainty (in percent over the true value) in the estimation of the source-frame chirp mass (y axis) against the true source-frame chirp mass (x axis). The colorbar is the redshift of the sources. A star reports the coordinates of \TheEvent{}. }
\label{Fig.5IFO_GammaMCS_MCD}
\end{figure}

Let us start with the relative uncertainties in the source-frame chirp mass, Fig.~\ref{Fig.5IFO_GammaMCS_MCD}. Comparing with the corresponding plot for the HLV network, Fig.~\ref{Fig.GammaMCS_MCD}, we see that uncertainties are similar, and mostly around $\sim30\%$. The bulk of the distribution is slightly larger for HLVIJ because more events with high redshifted mass are detected by this network, owing to its larger range.

\begin{figure}[htb]
\includegraphics[width=0.95\columnwidth]{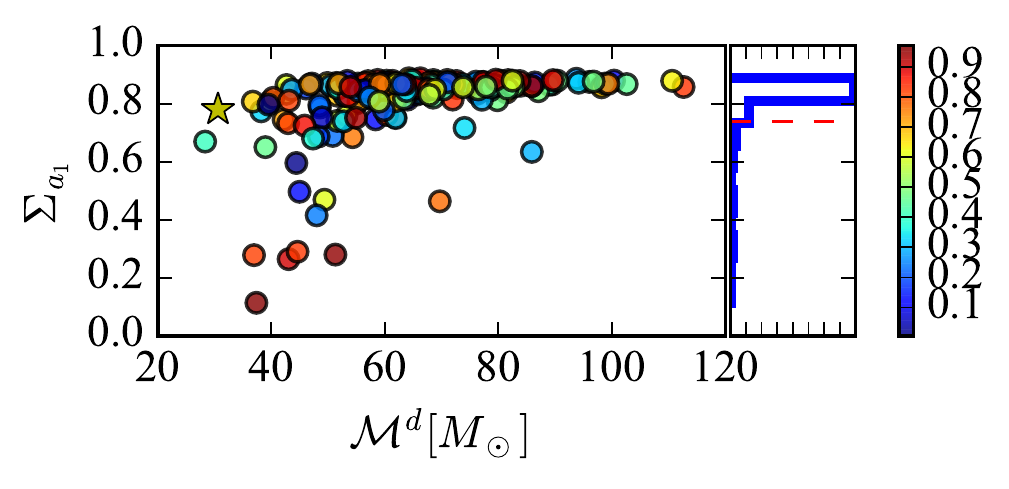}
\caption{For an HLVIJ network, the distribution of the 90\% CI uncertainty in the estimation of the primary spin magnitude (y axis) against the true detector-frame chirp mass (x axis). The colorbar shows the magnitude of the primary spin. A star reports the coordinates of \TheEvent{}.}
\label{Fig.5IFO_SigmaA1_MC}
\end{figure}

We will not plot the distribution of uncertainties for $m_1$ and $m_2$, but just mention they too look very similar to the corresponding HLV curves. In particular, the relative source-frame $m_1$ ($m_2$) uncertainty peaks at $\sim45$\% ($\sim50$\%), which is slightly more than for HLV, for the reasons just mentioned above.
In Fig.~\ref{Fig.5IFO_SigmaA1_MC} we show instead the uncertainties for the spin magnitude of the primary. We still find that large errors will be common, with only 10\% of the systems having 90\% CI below 0.73 (basically the same as HLV, for which we obtained 0.70). Once again, measurement is harder for the spin of the secondary object, 90\% of the sources will have uncertainties above 0.86, i.e. they will be unmeasurable. 

Fig.~\ref{Fig.5IFO_SigmaA1_MC} also shows that measurement of spins gets worse for systems with large (redshifted) mass. We have seen above, Sec.~\ref{Sec.DepOnMass} how this is indeed the case.

\begin{figure}[htb]
\includegraphics[width=0.95\columnwidth]{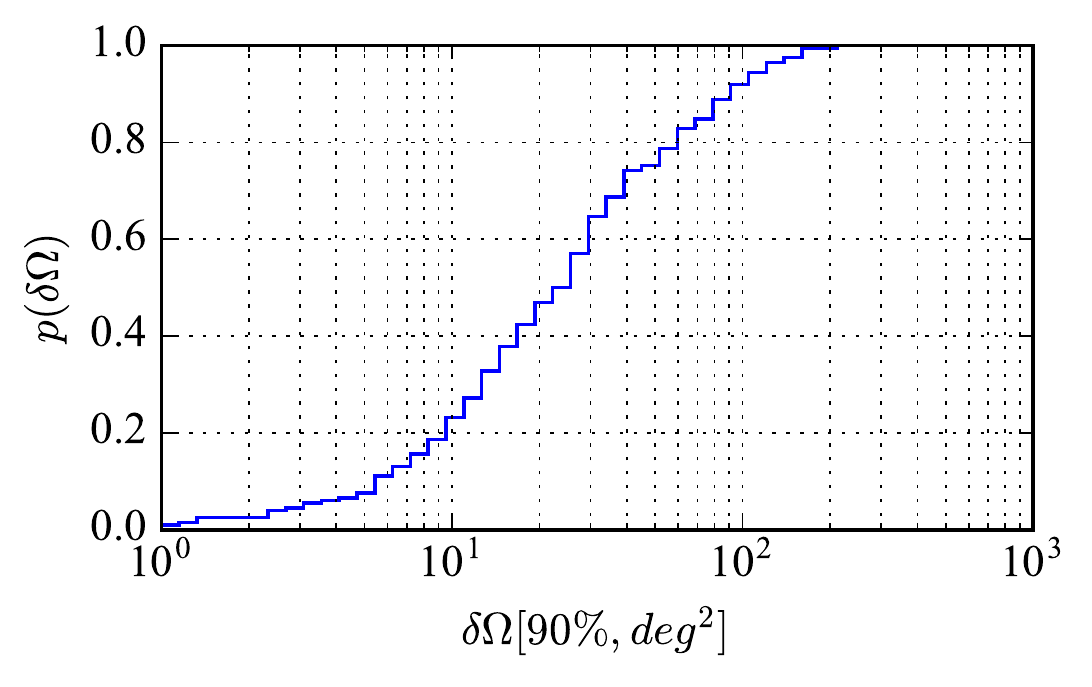}
\caption{Cumulative distribution for the 90\% CI sky localization area for BBH detected by the HLVIJ network.}
\label{Fig.5IFO_Sky}
\end{figure}

We end this appendix by mentioning that, as one would expect, sky localization gets better with the 5-detector network. Using the same figure of merit of Sec.~\ref{Sec.DistSky} we find that the median sky localization uncertainty is \si$25$~deg$^2$, i.e. a factor of \si~2 smaller than what obtained with the HLV network, Fig.~\ref{Fig.5IFO_Sky}.

\bibliography{pe}

\end{document}